\documentclass[a4paper,11pt]{article}
\pdfoutput=1 

\usepackage{jheppub_VVK} 

\usepackage[T1]{fontenc} 

\usepackage[utf8]{inputenc} 

\usepackage{caption} 
\usepackage{graphicx}
\usepackage{amssymb}

\def\sst{\scriptscriptstyle}

\usepackage{tikz}
\usetikzlibrary{trees}

\usetikzlibrary{trees}

\usetikzlibrary{decorations.pathmorphing}
\usetikzlibrary{decorations.markings}

	\tikzset{
	  photon/.style={decorate, decoration={snake}, draw=npurple,very thick},
	  boson/.style={decorate, decoration={snake}, draw=black, thick},	  electron/.style={draw=black,very thick, postaction={decorate},
	           decoration={markings,mark=at position .55 with 
{\arrow[draw=black]{>}}}
	  },
	  fermion/.style={draw=jblue,very thick, postaction={decorate},
	            decoration={markings,mark=at position .55 with 
{\arrow[draw=jblue]{}}}
	  },
	  higgs/.style={draw=wred,very thick, postaction={decorate},
	           decoration={markings,mark=at position .55 with 
{\arrow[draw=wred]{>}}}
	  },
	  nothing/.style={draw=white,very thick}
	  }

\usepackage{natbib}
\preprint{IPPP/16/43, DCPT/16/86}

\title{\boldmath Dark Matter and Leptogenesis Linked by  Classical Scale Invariance}

\author{Valentin V. Khoze}
\author{and Alexis D. Plascencia}

\affiliation[]{Institute for Particle Physics Phenomenology, Department of Physics,\\
Durham University, Durham DH1 3LE, United Kingdom}

\emailAdd{valya.khoze@durham.ac.uk}
\emailAdd{a.d.plascencia-contreras@durham.ac.uk}

\abstract{In this work we study a classically scale invariant extension of the Standard Model that can explain simultaneously 
dark matter and the baryon asymmetry in the universe. In our set-up we introduce a dark sector, namely a 
non-Abelian SU(2) hidden sector coupled to the SM via the Higgs portal, and a singlet sector 
responsible for generating Majorana masses for three right-handed sterile neutrinos. The gauge bosons
of the dark sector are mass-degenerate and stable, and this makes them suitable as 
dark matter candidates. Our model also accounts for the matter-anti-matter asymmetry.
The lepton flavour asymmetry is produced during CP-violating oscillations of the GeV-scale right-handed neutrinos, and converted
to the baryon asymmetry by the electroweak sphalerons. All the characteristic scales in the model: the electro-weak, 
dark matter and the leptogenesis/neutrino mass scales, are generated radiatively, have a common origin and related to each other 
via scalar field couplings in perturbation theory.}

\begin{document} 
\maketitle
\flushbottom

\section{Introduction}
\label{sec:intro}

The question of why the only scale parameter in the Standard Model (SM) Lagrangian, $-M^2_{\text{\tiny SM}} |H|^2$,
is much smaller than the Planck scale is at heart of the naturalness problem.
The idea of generating a scale radiatively, originally proposed in Ref.~\cite{Coleman:1973jx} can be applied to explain the origin of the electroweak scale in the SM \cite{Bardeen:1995kv, Hempfling:1996ht}.
In this article we will discuss an extension of the Standard Model that addresses some of the main shortcomings of the minimal theory, namely
the dark matter (DM), the baryon asymmetry of the universe (BAU) and the origin of the electroweak scale. 
Our Beyond the Standard Model framework is based on a theory which contains no explicit mass-scale parameters in its
tree-level Lagrangian, and all new scales will be generated dynamically at or below the TeV scale.
Our specific approach is motivated by the earlier work in 
Refs.~\cite{Chang:2007ki,Englert:2013gz,Khoze:2013oga,Hambye:2013sna,Carone:2013wla,Khoze:2014xha,Karam:2015jta} and 
\cite{Akhmedov:1998qx,Drewes:2012ma}.
The idea of generating the electro-weak scale and various scales of new physics via quantum corrections, 
by starting from a classically scale-invariant theory, has generated a lot of interest. For related studies on this subject 
we refer the reader to recent papers including
Refs.~\cite{Meissner:2006zh, Foot:2007ay, Foot:2007iy, Iso:2009ss, Holthausen:2009uc, AlexanderNunneley:2010nw, Lee:2012jn, Heikinheimo:2013fta, Farzinnia:2013pga, Khoze:2013uia,  Gabrielli:2013hma, Tamarit:2013vda,  Abel:2013mya, Allison:2014zya, 
Benic:2014aga, Plascencia:2015xwa, Ghorbani:2015xvz, Ahriche:2016ixu}.

In our set-up we extend the Standard Model by a dark sector, namely a 
non-Abelian SU(2$)_{\text{DM}}$ hidden sector that is coupled to the Standard Model via the 
Higgs portal, and a singlet sector that includes a real singlet $\sigma$ and 
three right-handed Majorana neutrinos $N_i$. Due to an SO(3) custodial symmetry all 
three gauge bosons $Z'^{\hspace{0.25mm}a}$ have the same mass and are absolutely stable, making 
them suitable dark matter candidates \cite{Hambye:2008bq} (this also applies to larger gauge groups 
SU(N$)_{\text{DM}}$ \cite{DiChiara:2015bua, Gross:2015cwa} and to scalar fields 
in higher representations \cite{Khoze:2014woa},
albeit symmetry breaking patterns get more complicated). 

The tree-level scalar potential of our model is given by
%
\begin{eqnarray}
V_0  & =& \lambda_{\phi} |\Phi|^4 + \lambda_h 
|H|^4 + \frac{\lambda_{\sigma}}{4} \sigma^4 - \lambda_{h\phi} 
 |H|^2 |\Phi|^2  - \frac{\lambda_{\phi \sigma}}{2} 
 |\Phi|^2 \sigma^2  + \frac{\lambda_{h \sigma}}{2}|H|^2
\sigma^2,  
\label{eq:CSIPotential}  
\end{eqnarray}
%
where $\Phi$ denotes the SU(2$)_{\text{DM}}$ doublet, $H$ is the SM Higgs doublet, and $\sigma$ is a gauge-singlet 
introduced in order to generate the Majorana masses for the sterile neutrinos, and hence the visible neutrinos masses and mixings 
via the see-saw mechanism. 
The portal couplings $\lambda_{h\phi}$, $\lambda_{\phi\sigma}$ and $\lambda_{h\sigma}$ will play a role in
order to induce non-trivial vacuum expectation values for all three scalar. 
As will become clear 
from Table~\ref{tab:ranges} 
we will scan over positive as well as negative values of the portal couplings $\lambda_{h\phi}$ and $\lambda_{h\sigma}$.
As we are working with multiple scalars we will adopt the 
Gildener-Weinberg approach \cite{Gildener:1976ih}, which is a generalisation of the Coleman-Weinberg  mechanism to 
multiple scalar states and will be briefly reviewed in Section {\bf \ref{sec:2}}.
Later on we shall see that
the most interesting region in parameter space leading to both the correct dark matter abundance
and the correct baryon asymmetry is for $\langle \sigma \rangle \gg \langle \phi \rangle$ and
hence one can think of $\sigma$ as a Coleman-Weinberg
scalar that once it acquires a non-zero vev it will be communicated to $\phi$ and 
$h$ through the portal couplings $\lambda_{\phi \sigma}$ and $\lambda_{h \sigma}$.

The interactions for the right-handed neutrinos in the Lagrangian are given by 
%
\begin{equation}
\label{eq:LagN}
\mathcal{L}_N = -\frac{1}{2} \left( Y_{ij}^M \sigma \overline{N_i}^c N_j + 
Y_{ij}^{M\dagger} \sigma \overline{N_i} N_j^c \right) - Y_{ia}^D \hspace{0.5mm} 
\overline{N_i} (\varepsilon H) l_{La} - Y_{ai}^{D\dagger} \hspace{0.5mm} 
\overline{l_{La}} (\varepsilon H)^{\dagger} N_i,
\end{equation}
%
where the first two term give rise to the Majorana mass once $\sigma$ acquires 
a vev, while the last two terms are responsible for the CP-violating 
oscillations of $N_i$.

Since we do not wish to break the lepton-number symmetry explicitly, it follows from \eqref{eq:LagN} 
that our new singlet scalar field $\sigma$ should have the lepton number $L=-2$. We can think of it as 
the real part of a complex scalar $\Sigma=( \sigma+i\pi)/\sqrt{2}$ where $S$ transforms under a $U(1)_L$ symmetry associated with the lepton number, which is broken spontaneously by $\langle \sigma\rangle \neq 0$.
If this is a global $U(1)$ symmetry then there must exist a massless (or very light) (pseudo)-Goldstone boson. Since the Higgs can pair-produce them and decay, this would severely constrain the portal coupling  of $\Sigma$ with the Higgs, $\lambda_{h \sigma} < 10^{-5}$,
see e.g. Ref.~\cite{Englert:2013gz}.
If we wish to avoid such fine-tuning, a much more appealing option would be to gauge the lepton number. A compelling scenario
is the $B\!-\!L$ theory with the anomaly free $U(1)_{B-L}$ factor. The generation of matter-anti-matter asymmetry via a leptogenesis
mechanism through sterile neutrino oscillations in a classically-scale-invariant $U(1)_{B-L} \times {\rm SM}$ theory was considered in Ref.~\cite{Khoze:2013oga}, and their results will also apply to our model. The main difference with the set-up followed in this paper is that here
we allow for a separate non-Abelian Coleman-Weinberg sector (i.e. it remains distinct from the $U(1)_{B-L}$ gauge sector) and as a result
we have a non-Abelian vector DM candidate.

Finally, it should also be possible to restrict the complex singlet $\Sigma$ back to the real singlet $\sigma$, just as we have in
\eqref{eq:CSIPotential}. In this case the continuous lepton number $U(1)$ symmetry is reduced to a discrete sub-group:
\begin{equation}
\sigma \to - \sigma\,, \quad (N,  \overline{N}^c, l_{L})\, \to \,e^{i\pi/2} (N,  \overline{N}^c, l_{L})\,, \quad
(\overline{N},N^c,\overline{l_L}) \,\to \, e^{-i\pi/2} (\overline{N},N^c,\overline{l_L})\,.
\label{eq:discr}
\end{equation}
In general all three possibilities corresponding to global, local and discrete lepton-number symmetries can be accommodated
and considered simultaneously in the context of Eqs.~\eqref{eq:CSIPotential}-\eqref{eq:LagN}
by either working with the real scalar $\sigma$ or the complex one by promoting $\sigma \to \sqrt{2} \Sigma$ 
(or $ \sqrt{2} \Sigma^\dagger$ in the second term in the brackets on the r.h.s. of \eqref{eq:LagN}). In this work we consider $\sigma$ to be a real scalar singlet.

\medskip

\section{From Coleman-Weinberg to the Gildener-Weinberg mechanism}
\label{sec:2}

The scalar field content of our model consists of an SU(2$)_L$ doublet $H$, an SU(2$)_{\text{DM}}$ doublet $\Phi$ and a real scalar $\sigma$; the latter giving mass to the sterile neutrinos after acquiring a vev in similar fashion to Ref.~\cite{Karam:2015jta}. Working in the unitary gauge of the SU(2$)_L\times$SU(2$)_{\text{DM}}$, the two scalar doublets in the theory are reduced to,
\begin{equation}\nonumber
H =  \frac{1}{\sqrt{2}} \begin{pmatrix}
  0 \\ h
 \end{pmatrix}, \hspace{12mm} 
 \Phi = \frac{1}{\sqrt{2}} \begin{pmatrix}
  0 \\ \phi
 \end{pmatrix}\,,
\end{equation}
and the tree-level potential becomes,
\begin{equation}\label{eq:3scsV}
V_0 = \frac{\lambda_h}{4} h^4 + \frac{\lambda_{\phi}}{4} \phi^4  + \frac{\lambda_{\sigma}}{4} \sigma^4 - \frac{\lambda_{h\phi}}{4} h^2 \phi^2  - \frac{\lambda_{\phi\sigma}}{4} \phi^2 \sigma^2  + \frac{\lambda_{h\sigma}}{4} h^2 \sigma^2\,.
\end{equation}
There are no mass scales appearing in the classical theory, and at the origin in the field space,
all scalar vevs are zero, in agreement with classical scale invariance. We impose a conservative constraint on all the scalar couplings for the model to be perturbative $| \lambda_i|\!<\!3$,
we also impose $g_{\text{\tiny{DM}}}\!<\!3$ and in order to ensure vacuum stability the following set of constraints need to be satisfied,
\begin{eqnarray}\label{eq:Stability1} 
\lambda_h\geq 0, \hspace{15mm}   \lambda_{\phi}\geq 0, \hspace{13mm}   \lambda_{\sigma}\geq 0 , \hspace{12mm}  \\[1ex]
\frac{\lambda_{h\phi}}{2\sqrt{\lambda_h\lambda_{\phi}}} \leq 1, \hspace{2mm}  
 -\frac{\lambda_{h\sigma}}{2\sqrt{\lambda_h\lambda_{\sigma}}} \leq 1, \hspace{2mm} 
  \frac{\lambda_{\phi\sigma}}{2\sqrt{\lambda_{\phi}\lambda_{\sigma}}} \leq 1, \hspace{2mm}  \\[1ex]
\frac{\lambda_{h\phi}}{2\sqrt{\lambda_h\lambda_{\phi}}}\,-\frac{{\lambda_{h\sigma}}}{2\sqrt{\lambda_h\lambda_{\sigma}}}\,+\,\frac{\lambda_{\phi\sigma}}{2\sqrt{\lambda_{\phi}\lambda_{\sigma}}}\,\leq\,1. \hspace{15mm} 
\end{eqnarray}
For more detail we refer to Table~\ref{tab:ranges}.

\medskip

\subsection{The Coleman-Weinberg approximation}

For simplicity, let us temporarily ignore the singlet $\sigma$ and concentrate on the theory
with two scalars, $\phi$ and $h$. We will further refer to the hidden SU(2$)_{\text{DM}}$ sector
with $\phi$ as the Coleman-Weinberg (CW) sector.
In the near-decoupling limit, $\lambda_{h\phi}\ll 1$, between the CW and the SM sectors, we can view
electroweak symmetry breaking effectively as a two-step process \cite{Englert:2013gz}.

First, the CW mechanism \cite{Coleman:1973jx} generates 
$\langle \phi \rangle$ in the hidden sector through running couplings (or more precisely the dimensional transmutation).  
To make this work, 
the scalar self-coupling $ \lambda_{\phi}$ at the relevant scale $\mu=\langle \phi \rangle$ 
should be small -- of the order of $g_{\text{\tiny{DM}}}^4 \ll 1$, as we will see momentarily. 
This has the following interpretation: in a theory where $\lambda_\phi$ has a positive slope, we start at a relatively high scale where 
$\lambda_\phi$ is positive and move toward the infrared until approach the value of the
$\mu$ where $\lambda_\phi (\mu)$ becomes small and is about to cross zero. This is the Coleman-Weinberg scale where
the potential develops a non-trivial minimum and $\phi$ generates a non-vanishing vev. 

To see this, consider the 1-loop effective potential evaluated at the scale $\mu$ ({\it cf.} \cite{Khoze:2014xha}):
\begin{equation}
V(\phi,h)=\frac{\lambda_\phi(\mu)}{4}\phi^4 +\frac{9 }{1024\,\pi^2}\, g^4_{\sst \mathrm{DM}}(\mu)
\,\phi^4\left(\log\frac{\phi^2}{\mu^2}-\frac{25}{6}\right)
-\frac{\lambda_{h\phi}(\mu)}{4} h^2 \phi^2\,,
\end{equation}
Here we are keeping 1-loop corrections arising from interactions of
$\phi$ with the SU(2) gauge bosons in the hidden sector, but neglecting the loops of $\phi$ (since $\lambda_{\phi}$ is close to zero)
and the radiative corrections from the Standard Model sector. The latter would
produce only subleading corrections to the vevs.  Minimising at $\mu=\left<\phi\right>$ gives:
\begin{equation}
\lambda_\phi \,= \, \frac{33}{256\,\pi^2} \,g_{\sst \mathrm{DM}}^4 
+\lambda_{h\phi}\frac{v^2}{2\langle\phi\rangle^2}
\qquad {\rm at} \quad \mu=\langle \phi\rangle\,.
\label{eq:cwmsbar-PSU2}
\end{equation}
For small portal coupling $\lambda_{h\phi}$, this is a small deformation of the
original CW condition, $\lambda_{\phi}(\langle\phi\rangle) =\, \frac{33}{256\,\pi^2} \,g_{\sst \mathrm{DM}}^4 (\langle \phi\rangle)$.

The second step of the process is the transmission of the vev $\langle \phi  \rangle$  to the
Standard Model via the Higgs portal, generating a negative mass squared 
parameter for the Higgs $=-\lambda_{h\phi}\langle \phi^{2}\rangle$
which generates the
electroweak scale $v$,
\begin{equation}
v = \langle h \rangle  \, =\, 
  \sqrt{\frac{2\lambda_{h\phi}}{\lambda_h}} \, 
   \langle \phi\rangle \,, \quad m_h \,=\, \sqrt{2 \lambda_h} v\,.
   \label{eq:v}
\end{equation}
The fact that for $\lambda_{h\phi}\ll 1$ the generated electroweak scale is much smaller than
$\langle  \phi  \rangle$, guarantees that any back reaction on the hidden
sector vev $\langle \phi \rangle$ is negligible.
Finally, the mass of the CW scalar is obtained from the 1-loop potential and reads:
\begin{equation}
m_\phi^2 \,=\,
\frac{9 }{128\,\pi^2}\,g_{\sst \mathrm{DM}}^4\, \langle \phi \rangle^2
+\lambda_{h\phi} v^2\,.
\label{eq:mphiZ-PSU2}
\end{equation}

As already stated, this approach is valid in the near-decoupling approximation where all the portal couplings are small.
The dynamical generation of all scales is visualised here as first the generation of the CW scalar vev $\langle \phi\rangle$,
which then induces the vevs of other scalars proportional to the square root of the corresponding portal couplings $\ll 1$, as in
\eqref{eq:v}. This implies the hierarchy of the vevs.

For multiple scalars, $\phi$, $h$ and $\sigma$, it is not a priori obvious why the portal couplings should be small and
which of the scalar vevs should be dominant.
For example on one part of the parameter space we can find $\langle \phi \rangle > \langle \sigma\rangle$ and 
on a different part one has $ \langle \sigma\rangle > \langle \phi \rangle$ (so that $\sigma$ rather than $\phi$ effectively
plays the role of the CW scalar). To consider all such cases and not be constrained by the near-decoupling limits we will
utilise the Gildener-Weinberg set-up \cite{Gildener:1976ih}, which is a generalization of the Coleman-Weinberg  method.

\subsection{The Gildener-Weinberg approach}

We now return to the general case with the three scalars in the model are described by the tree-level massless scalar 
potential \eqref{eq:3scsV}. The Gildener-Weinberg mechanism was recently worked out for this case in Ref.~\cite{Karam:2015jta},
which we will follow.
All three vevs can be generated dynamically but neither of the original scalars is solely
responsible for the intrinsic scale generation; this instead is a collective effect generated by a linear combination of all three
scalars $\varphi$.

Following \cite{Gildener:1976ih}, we change variables and reparametrise the scalar fields via,
\begin{eqnarray}\label{eq:ScalarParam} 
h= N_1 \varphi,  \hspace{5mm} &     \phi= N_2 \varphi , \hspace{5mm}  &     \sigma=N_3 \varphi . 
\end{eqnarray}
where each $N_i$ is a unit vector in three-dimensions. The Gildener-Weinberg mechanism tells us that a non-zero vacuum expectation value will be generated in some direction in scalar field space $N_i\!=\!n_i$,  so this direction must satisfy the condition,
\begin{equation}\label{eq:Minimum} 
\left. \frac{\partial V_0}{\partial N_i} \right|_\mathbf{n} = 0 ,
\end{equation}
and furthermore the value of the tree-level potential in this vacuum is independent of $\varphi$,
\begin{equation}
\label{eq:ScalarParam2} 
V_0 (n_1\varphi , n_2\varphi , n_3\varphi ) = 0\,.
\end{equation}
The latter condition is simply the statement that the potential restricted to the single degree of freedom $\varphi$, 
is of the form $\frac{1}{4}\lambda_{\varphi} \, \varphi^4$ with the corresponding coupling constant vanishing $\lambda_{\varphi}=0$.
This is nothing but the definition of scale $\mu_{\text{\tiny{GW}}}$ where $\lambda_{\varphi}(\mu_{\text{\tiny{GW}}})$ vanishes,
and is a reflection of a similar statement in the Coleman-Weinberg case for the single scalar that its self-coupling 
was about to cross zero, but was stabilised at the small positive value by the gauge coupling
at the Coleman-Weinberg scale $\mu_{\text{\tiny{CW}}}$, see Eq.~\eqref{eq:cwmsbar-PSU2}.

Being a unit vector in three-dimensions, $n_i$'s can be parametrised in terms of two independent angles, $\alpha$ and $\gamma$
and we will call the $\varphi$ vev, $w$, so that,
\begin{eqnarray}
\label{eq:ns}
n_1 &=& \sin\alpha\,, \quad n_2=\cos\alpha \cos \gamma\,\quad n_3 = \cos\alpha \sin \gamma\,,
\\
\label{eq:wvevs}
\langle h \rangle &=& w n_1\,, \quad  \langle \phi \rangle = w n_2\,, \quad \qquad  \langle \sigma \rangle = w n_3.
\end{eqnarray}
The three linearly-independent conditions arising from the Gildener-Weinberg minimisation \eqref{eq:Minimum} of the tree-level potential
amount to the following set of relations,
\begin{eqnarray}
2 \lambda_h n_1^2 \hspace{1mm} &  = & \lambda_{h\phi} n_2^2 -  \lambda_{h\sigma}n_3^2,\label{eq:GW1} \\[1ex]
2 \lambda_{\phi} n_2^2 \hspace{1mm} &  = & \lambda_{h\phi} n_1^2 +  \lambda_{\phi\sigma}n_3^2,\label{eq:GW2} \\[1ex]
2 \lambda_{\sigma} n_3^2 \hspace{1mm} &  = & \lambda_{\phi\sigma} n_2^2 -  \lambda_{h\sigma}n_1^2.
\label{eq:GW3} 
\end{eqnarray}
These conditions hold at the scale $\mu_{\text{\tiny{GW}}}$ where the scalar fields develop the vev $\langle \varphi \rangle =w$ \eqref{eq:wvevs}.
Due to the three scalars acquiring non-zero vacuum expectation values, the three states will mix among each other. 
The mass matrix $M^2$ is diagonalised for $h_1, h_2$ and $h_3$ eigenstates via the rotation matrix $O$,
\begin{equation}
\label{eq:eigen}
{\rm diag} \left(M^2_{h_1},M^2_{h_2},M^2_{h_3}\right)\,=\, O^{(-1)}\, M^2 \, O\,, \qquad
\begin{pmatrix} h \\ \phi \\ \sigma \end{pmatrix}  = O_{ij} \begin{pmatrix} h_1 \\ h_2 \\ h_3 \end{pmatrix}, 
\end{equation}
%
and we further identify the state $h_1$ with the SM 125 GeV Higgs boson. Following \cite{Karam:2015jta} we parametrise the 
rotation matrix in terms of
three mixing angles $\alpha$, $\beta$ and $\gamma$,
\begin{equation}\label{eq:rotationmatrix} 
O =  \begin{pmatrix}
  \cos \alpha \cos \beta & \sin \alpha & \cos \alpha \sin \beta  \\ 
  -\cos \beta \cos \gamma \sin \alpha + \sin \beta \sin \gamma  \hspace{3mm} &  \cos \alpha \cos \gamma \hspace{3mm} & -\cos \gamma \sin \alpha \sin \beta - \cos \beta \sin \gamma \\ 
  -\cos \gamma \sin \beta - \cos \beta \sin \alpha \sin \gamma \hspace{3mm} & \cos \alpha \sin \gamma \hspace{3mm} & \cos \beta \cos \gamma - \sin \alpha \sin \beta \sin \gamma
 \end{pmatrix},
\end{equation}
and use it to compute the scalar mass eigenstates \eqref{eq:eigen} at tree-level. The resulting
expressions for the scalar masses can be found in Ref.~\cite{Karam:2015jta}. 
There is one classically flat direction in the model -- along $\varphi$ -- in which the potential develops the vacuum expectation value.
Our choice of parametrisation in \eqref{eq:wvevs} and in the second row of \eqref{eq:rotationmatrix} in terms of the same two angles
$\alpha$ and $\gamma$, selects this direction to be identified with $h_2$. Hence, at tree level, $M_{h_2}=0$, but it will become
non-zero, see Eq.~\eqref{eq:mh2} below, when one-loop effects are included.

At the scale $\mu_{\text{\tiny{GW}}}$ the one-loop effective potential along the minimum flat direction can be written 
as \cite{Gildener:1976ih},
\begin{equation}\label{eq:1} 
V(\varphi \mathbf{n}) = A\varphi^4 + B \varphi^4 \log \left( \frac{\varphi^2}{\mu_{\text{\tiny{GW}}}^2} \right),
\end{equation}
where the $A$ and $B$ coefficients are computed in the $\overline{\text{MS}}$ \cite{Martin:2001vx} scheme and given by,
\begin{eqnarray}\label{eq:ABcoefficients} 
A & = & \frac{1}{64\pi^2 w^4} \left[ \sum_{i=1,3} M^4_{h_i} \left( -\frac{3}{2} + \log \frac{M^2_{h_i}}{w^2} \right) +6 M^4_W \left( -\frac{5}{6} + \log \frac{M^2_W}{w^2}  \right) + 3 M^4_Z \left( -\frac{5}{6} + \log \frac{M^2_Z}{w^2} \right) \right.\nonumber\\
 & & \left. + \hspace{1mm} 9 M^4_{Z'} \left( -\frac{5}{6} + \log \frac{M^2_{Z'}}{w^2}  \right) - 12 M^4_t \left( -1 + \log \frac{M^2_t}{w^2}  \right) - 2 \sum^3_{i=1} M^4_{N_i} \left( -1 + \log \frac{M^2_{N_i}}{w^2}  \right) \right], \nonumber\\[1ex]
B & = & \frac{1}{64\pi^2 w^4} \left(\sum_{i=1,3} M^4_{h_i}  +6 M^4_W  + 3 M^4_Z 
  +  9 M^4_{Z'}  - 12 M^4_t  - 2 \sum^3_{i=1} M^4_{N_i}  \right)\,, \nonumber
\end{eqnarray}
where $M_{h_i}$ are the tree-level masses of the three scalar eigenstates, $h_1$, $h_2$ and $h_3$, and the rest are the 
masses of the SM and the hidden sector vector bosons as well as the top quark and the right-handed Majorana neutrinos.
We can now see that at the RG scale $\mu_{\text{\tiny{GW}}}$ the 1-loop corrected effective
potential has a fixed vacuum expectation value $w$ that satisfies,
\begin{equation}\label{eq:2} 
\log \left( \frac{w}{\mu_{\text{\tiny{GW}}}} \right)  = - \frac{1}{4} - \frac{A}{2B}\, ,
\end{equation}
and using this relation we can rewrite the one-loop effective potential as,
\begin{equation}\label{eq:3} 
V = B \varphi^4 \left( \log \frac{\varphi^2}{w^2} - \frac{1}{2} \right)\,,
\end{equation}
and we can also evaluate the potential at the minimum to be $V(\varphi\!=\!w)\!=\!-Bw^4/2$, which gives the requirement $B>0$ for this to be  a lower minimum than the one at the origin.
The mass of the pseudo-dilaton $h_2$ is then given by,
\begin{equation}
\label{eq:mh2}
M^2_{h_2}= \left. \frac{\partial^2 V}{\partial \varphi^2} \right|_\mathbf{n} = \frac{1}{8\pi w^2} \left( M_{h_1}^4 + M_{h_3}^4 + 6M_W^4 + 3M_Z^4  + 9M_{Z'}^4 - 12 M_t^4 - 2 \sum^3_{i=1} M^4_{N_i} \right).
\end{equation}
%

%
\begin{figure}[tbp]
  \centering				
				
				\begin{tikzpicture}[
					thick,
			level/.style={level distance=2.0cm, line width=0.4mm},
			level 2/.style={sibling angle=60},
			level 3/.style={sibling angle=60},
			level 4/.style={level distance=1.4cm, sibling angle=60}
	]
			
			\draw[boson] (-1.8,0) -- (-1,-1);
			\draw[boson]  (-1,-1) -- (-1.8,-2);
			\draw[black, dashed] (-1,-1) -- (0.6,-1);
		    \draw[boson] (0.6,-1) -- (1.4,0);
		    \draw[boson] (0.6,-1) -- (1.4,-2);

			\node[font=\normalsize] at (-2.1,0) {$Z'_a$};
			\node[font=\normalsize] at (-2.1,-2) {$Z'_a$};
			\node[font=\normalsize] at (-0.2,-.75) {$h_i$};
			\node[font=\normalsize] at (-2.1+4.2,0) {$Z, W^+$};
			\node[font=\normalsize] at (-2.1+4.2,-2) {$Z, W^-$};

			\draw[boson] (2.3+1.5,0) -- (3.1+1.5,-1);
			\draw[boson] (-1+5.6,-1) -- (-1.8+5.6,-2);
			\draw[black, dashed] (-1+5.6,-1) -- (0.6+5.6,-1);
		    \draw[electron] (6.2,-1) -- (7.1,0);
		    \draw[electron] (7.1,-2) -- (6.2,-1);

			\node[font=\normalsize] at (-2.1+5.6,0) {$Z'_a$};
			\node[font=\normalsize] at (-2.1+5.6,-2) {$Z'_a$};
			\node[font=\normalsize] at (-0.2+5.6,-.75) {$h_i$};
			\node[font=\normalsize] at (-2.1+4.2+5.6,0) {$f, N$};
			\node[font=\normalsize] at (-2.1+4.2+5.6,-2) {$\bar{f}, \bar{N}$};

		\end{tikzpicture}
		
  \caption{Dark matter annihilation diagrams into Standard Model gauge bosons and fermions, we also include annihilation into right-handed neutrinos.} 
  \label{fig:AnnSM}
 \end{figure}
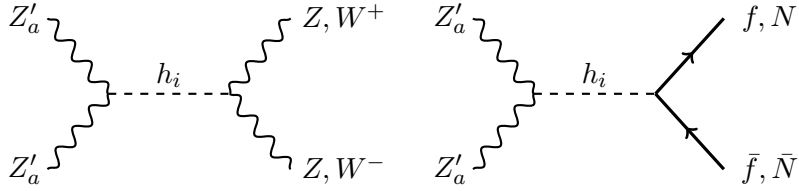
\begin{figure}[tbp]
  \centering				
				
				\begin{tikzpicture}[
					thick,
			level/.style={level distance=2.0cm, line width=0.4mm},
			level 2/.style={sibling angle=60},
			level 3/.style={sibling angle=60},
			level 4/.style={level distance=1.4cm, sibling angle=60}
	]
			
			\draw[boson] (-1.6,0) -- (-1,-1);
			\draw[boson]  (-1,-1) -- (-1.6,-2);
			\draw[black, dashed] (-1,-1) -- (-0.22,0);
			\draw[black, dashed] (-1,-1) -- (-0.22,-2);

			\node[font=\normalsize] at (-1.9,-0.2) {$Z'_a$};
			\node[font=\normalsize] at (-1.9,-1.8) {$Z'_a$};
			\node[font=\normalsize] at (0,-.18) {$h_i$};
			\node[font=\normalsize] at (0,-1.8) {$h_j$};

			\draw[boson] (2.3+1.5-2.5,0) -- (3.1+1.5-2.5,-1);
			\draw[boson] (-1+5.6-2.5,-1) -- (-1.8+5.6-2.5,-2);
			\draw[black, dashed] (-1+5.6-2.5,-1) -- (0.6+5.2-2.5,-1);
		    \draw[black, dashed] (5.8-2.5,-1) -- (6.8-2.5,0);
		    \draw[black, dashed] (5.8-2.5,-1) -- (6.8-2.5,-2);

			\node[font=\normalsize] at (-2.1+5.6-2.5,-0.2) {$Z'_a$};
			\node[font=\normalsize] at (-2.1+5.6-2.5,-1.8) {$Z'_a$};
			\node[font=\normalsize] at (-0.2+5.4-2.5,-.75) {$h_k$};
			\node[font=\normalsize] at (-2.1+4.2+5-2.6,-0.2) {$h_i$};
			\node[font=\normalsize] at (-2.1+4.2+5-2.6,-1.8) {$h_j$};

		    \draw[boson] (2.3+3.2+0.2,-0.2) -- (3.1+3.4+0.2,-0.2);
			\draw[boson] (-1.8+7.3+0.2,-1.9) -- (-1+7.5+0.2,-1.9);  
		    \draw[boson] (3.1+3.4+0.2,-0.2) -- (-1+7.5+0.2,-1.9);

		    \draw[black, dashed] (3.1+3.4+0.2,-0.2) -- (7.4+0.2,-0.2);
		    \draw[black, dashed] (-1+7.5+0.2,-1.9) -- (7.4+0.2,-1.9);
		    
			\node[font=\normalsize] at (5.3+0.2,-0.2) {$Z'_a$};
			\node[font=\normalsize] at (5.3+0.2,-1.8) {$Z'_a$};	
			\node[font=\normalsize] at (6.1+0.2,-1) {$Z'_a$};		    
			\node[font=\normalsize] at (-2.1+4.2+5.6+0.2,-0.2) {$h_i$};
			\node[font=\normalsize] at (-2.1+4.2+5.6+0.2,-1.9) {$h_j$};

		    \draw[boson] (2.3+3.2+0.2+3.2,-0.2) -- (3.1+3.4+0.2+3.2,-0.2);
			\draw[boson] (-1.8+7.3+0.2+3.2,-1.9) -- (-1+7.5+0.2+3.2,-1.9);  
		    \draw[boson] (3.1+3.4+0.2+3.2,-0.2) -- (-1+7.5+0.2+3.2,-1.9);

		    \draw[black, dashed] (3.1+3.4+0.2+3.2,-0.2) -- (7.4+0.2+3.3,-1.9);    
		    \draw[black, dashed] (-1+7.5+0.2+3.2,-1.9) -- (7.4+0.2+3.3,-0.2);
		    
			\node[font=\normalsize] at (5.3+0.2+3.2,-0.2) {$Z'_a$};
			\node[font=\normalsize] at (5.3+0.2+3.2,-1.8) {$Z'_a$};	
			\node[font=\normalsize] at (6.1+0.2+3.2,-1) {$Z'_a$};		    
			\node[font=\normalsize] at (-2.1+4.2+5.6+0.2+3.3,-0.2) {$h_i$};
			\node[font=\normalsize] at (-2.1+4.2+5.6+0.2+3.3,-1.9) {$h_j$};

		    

		\end{tikzpicture}
		
  \caption{Dark matter annihilation diagrams into scalar states.} 
  \label{fig:AnnScalars}
 \end{figure}
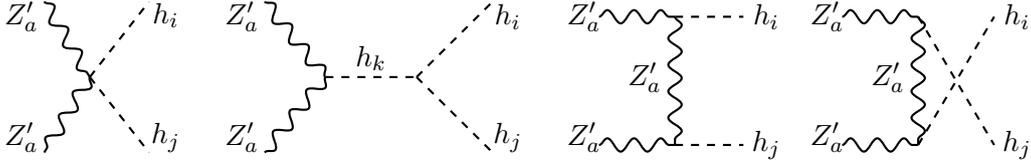
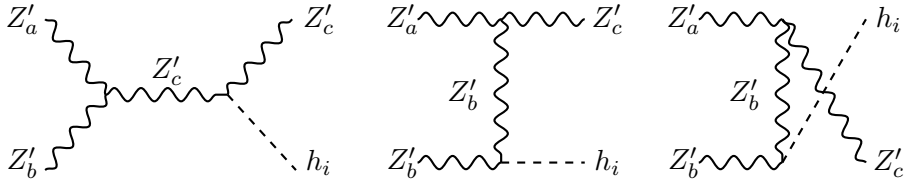
\begin{figure}[tbp]
  \centering				
				
				\begin{tikzpicture}[
					thick,
			level/.style={level distance=2.0cm, line width=0.4mm},
			level 2/.style={sibling angle=60},
			level 3/.style={sibling angle=60},
			level 4/.style={level distance=1.4cm, sibling angle=60}
	]
			
			\draw[boson] (-1.8,0) -- (-1,-1);
			\draw[boson]  (-1,-1) -- (-1.8,-2);
			\draw[boson] (-1,-1) -- (0.6,-1);
		    \draw[boson] (0.6,-1) -- (1.45,0);
		    \draw[black, dashed] (0.6,-1) -- (1.5,-2);

			\node[font=\normalsize] at (-2.1,0) {$Z'_a$};
			\node[font=\normalsize] at (-2.1,-1.9) {$Z'_b$};
			\node[font=\normalsize] at (-0.2,-.68) {$Z'_c$};
			\node[font=\normalsize] at (-2.1+3.9,0) {$Z'_c$};
			\node[font=\normalsize] at (-2.1+3.9,-1.9) {$h_i$};

		    \draw[boson] (-2.1+5.2, 0) -- (-2.1+5.2+1.1, 0);
			\draw[boson] (-2.1+5.2,-1.9) -- (-2.1+5.2+1.1,-1.9);  
		    \draw[boson] (-2.1+5.2+1.1,0) -- (-2.1+5.2+1.1,-1.9);

		    \draw[boson] (-2.1+5.2+1.1,0) -- (-2.1+5.2+2.2,0);
		    \draw[black, dashed] (-2.1+5.2+1.1,-1.9) -- (-2.1+5.2+2.2,-1.9);
		    
			\node[font=\normalsize] at (-2.1+5,0) {$Z'_a$};
			\node[font=\normalsize] at (-2.1+5,-1.9) {$Z'_b$};	
			\node[font=\normalsize] at (-2.1+5.2+0.6,-1) {$Z'_b$};	
		    \node[font=\normalsize] at (-2.1+5.2+2.2+0.3,0) {$Z'_c$};		    
			\node[font=\normalsize] at (-2.1+5.2+2.2+0.3,-1.9) {$h_i$};

		    \draw[boson] (-2.1+5.2+3.7, 0) -- (-2.1+5.2+1.1+3.7, 0);
			\draw[boson] (-2.1+5.2+3.7,-1.9) -- (-2.1+5.2+1.1+3.7,-1.9);  
		    \draw[boson] (-2.1+5.2+1.1+3.7,0) -- (-2.1+5.2+1.1+3.7,-1.9);

		    \draw[boson] (-2.1+5.2+2.2+3.7,-1.9)-- (-2.1+5.2+1.1+3.7,0);
		    \draw[black, dashed] (-2.1+5.2+1.1+3.7,-1.9) -- (-2.1+5.2+2.2+3.7,0);
		    
			\node[font=\normalsize] at (-2.1+5+3.7,0) {$Z'_a$};
			\node[font=\normalsize] at (-2.1+5+3.7,-1.9) {$Z'_b$};	
			\node[font=\normalsize] at (-2.1+5.2+0.6+3.7,-1) {$Z'_b$};	
		    \node[font=\normalsize] at (-2.1+5.2+2.2+0.3+3.7,0) {$h_i$};		    
			\node[font=\normalsize] at (-2.1+5.2+2.2+0.3+3.7,-1.9) {$Z'_c$};

		\end{tikzpicture}
		
  \caption{Vector dark matter semi-annihilation diagrams. In contrast to some other models of dark matter, $Z'_a$ is stable due to an remnant global symmetry.} 
  \label{fig:SemiDiagrams}
 \end{figure}

In summary, at the scale $\mu_{\text{\tiny{GW}}}$ the conditions Eqs.~\eqref{eq:GW1}--\eqref{eq:GW3} will be satisfied and the scalar potential will develop a non-trivial vev $w$ giving rise to non-zero vacuum expectation values $\langle h \rangle,$ $\langle \phi \rangle,$ and $\langle \sigma \rangle$. For one scalar field, the Coleman-Weinberg mechanism requires the scalar quartic coupling to take very small values $\lambda_{\phi} \sim g_{\text{\tiny{DM}}}^4$, in the Gildener-Weinberg scenario it is a combination of the quartic couplings that needs to vanish, so these couplings can take larger values individually.

The formulae for the mixing angles in terms of the coupling constants and the vevs follow from the diagonalisation
of the tree-level mass matrix,
\begin{eqnarray}\label{eq:MixAngles} 
\tan^2\alpha & = & \frac{\langle h \rangle^2}{\langle \phi \rangle ^2 + \langle \sigma \rangle ^2} = \frac{4\lambda_{\phi}\lambda_{\sigma}  -\lambda_{\phi \sigma}^2 }  {2(\lambda_{\sigma} \lambda_{h\phi} - \lambda_{\phi} \lambda_{h\sigma}) + \lambda_{\phi\sigma} (\lambda_{h\phi} - \lambda_{h\sigma}) }\,,\\[1.5ex]
\tan^2\gamma & = &   \frac{\langle \sigma \rangle ^2}{\langle \phi \rangle ^2 } = \frac{2\lambda_h\lambda_{\phi\sigma} -\lambda_{h\phi}\lambda_{h\sigma}  }         {4\lambda_h \lambda_{\sigma} - \lambda_{h\sigma}^2}\,, \\[1.5ex]
\tan2\beta & = & \frac{ \langle h \rangle \langle \phi \rangle \langle \sigma \rangle \hspace{0.2mm} w \hspace{0.2mm}  (\lambda_{h\sigma}+\lambda_{h\phi}) }{(\lambda_{\phi} + \lambda_{\sigma} + \lambda_{\phi \sigma}) \langle \phi \rangle ^2 \langle \sigma \rangle ^2 - \lambda_h \langle h \rangle^2 w^2}\,. 
\end{eqnarray}
Experimental searches of a scalar singlet mixing with the SM Higgs provide constraints on the mixing angles \cite{Martin-Lozano:2015dja, Robens:2015gla, Falkowski:2015iwa}. In our case, these translate as,
\begin{equation}
\label{anglelimit}
\cos ^2 \alpha \hspace{1mm} \cos^2 \beta > 0.85.
\end{equation}
In the region where the decay $h_1 \rightarrow h_2 h_2$ is allowed  we impose the stronger constraint $\cos ^2 \alpha \hspace{1mm} \cos^2 \beta > 0.96$. 
Nonetheless, due to the Gildener-Weinberg conditions the decay $h_1 \rightarrow h_2 \hspace{0.3mm} h_2$ is highly suppressed. 
In the scan we perform $M_{h_3}$ is always greater than $M_{h_1}$, so there is no need to worry about the SM Higgs decaying into two $h_3$ scalars. At the same time, strong constraints could come when the decays $h_1\rightarrow Z'^{\hspace{0.25mm}a} \hspace{0.3mm} Z'^{\hspace{0.3mm}a}$ are allowed, we set $M_{Z'}\!>\!M_{h_1}/2$ so that these decays are kinematically forbidden.

For the study of dark matter the Lagrangian contains ten dimensionless free parameters, which are reduced to eight
after fixing $\langle h \rangle\!=\!246$ GeV and $M_{h_1}\!=\!125$ GeV. We perform a
random scan on the remaining eight parameters in the ranges given in Table \ref{tab:ranges}.

\begin{table}[th!]
\centering
\begin{tabular}{|c| c|}
\hline
Parameter & Scan range \\ \hline
$\lambda_{\phi\sigma}$         & (0, 0.5)  \\
$\lambda_{h\phi}$              & (-0.5, 0.5)   \\
$\lambda_{h\sigma}   $         & (-0.25, 0.25)  \\
$\lambda_{\phi}$               & (0, 3)  \\
$g_{\text{\tiny{DM}}}$         & (0, 3)  \\
$M_{N_i}$                      & (0, 100) GeV \\
\hline
\end{tabular}
\caption{
  \label{tab:ranges}
Ranges for the input parameters in the scan.
}
\end{table}

The matrix $Y^D$ has no impact on the dark matter phenomenology, but it is crucial 
for Leptogenesis and it will be parametrised by three complex angles $\omega_{ij}$ using the Casas-Ibarra parametrisation \cite{Casas:2001sr}. Therefore, once we set all the parameters for the active neutrinos to their best experimental fit, there are twelve free parameters in the model.

\section{Dark matter phenomenology}
\label{sec:DM}

Evidence from astrophysics suggests that most of the matter in the universe is made out of cosmologically stable 
dark matter that interacts very weakly
with ordinary matter. Being able to identify what constitutes this dark matter is one of the deepest mysteries in both particle physics and
astrophysics. In this work we consider the possibility of dark matter being a spin-1 particle from a hidden sector with non-Abelian 
SU(2$)_{\text{DM}}$ gauged symmetry. The idea of vector dark matter was first introduced in Ref.~\cite{Hambye:2008bq} and later 
studied in Refs.~\cite{Hambye:2013sna, Boehm:2014bia, Khoze:2014xha, Gross:2015cwa}. Note that if the hidden sector had been U(1), 
the kinetic mixing among the hidden sector and the hypercharge will have made our dark matter candidate unstable.

\begin{figure}[tbp]
\centerline\\  \center
\scalebox{0.5}{\includegraphics{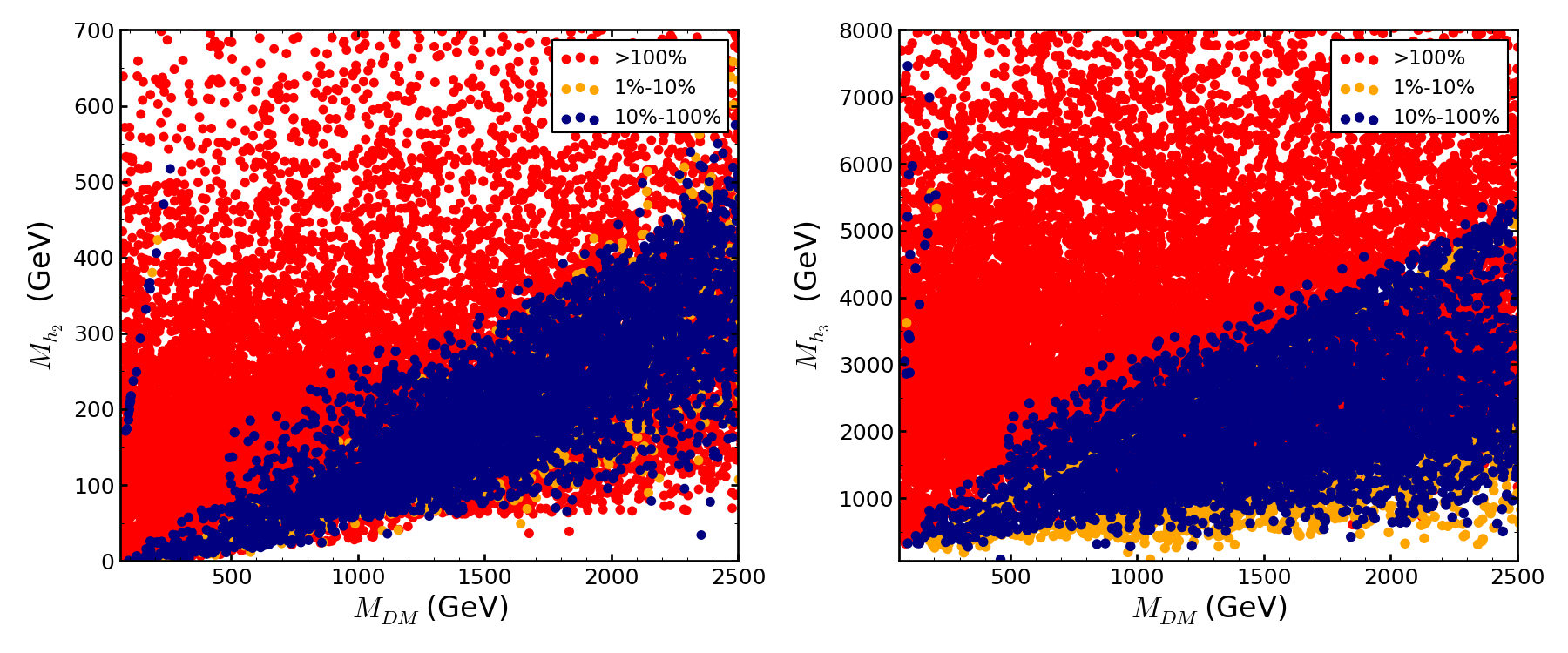}}
\caption{\label{fig:Masses} Left panel shows scatter plot of the dark matter mass $M_{\text{DM}}\!=\!M_{Z'}$ versus 
the scalar mass $M_{h_2}$. Right panel gives scatter plot of the dark matter mass versus the mass of the heavier scalar $h_3$.
Different colours indicate whether the vector gauge 
triplet accounts for more or less than 100$\%$, 10$\%$ and 1$\%$ of the observed dark matter 
abundance.}
\end{figure}
%

\begin{figure}[tbp]
\centerline\\  \center
\scalebox{0.5}{\includegraphics{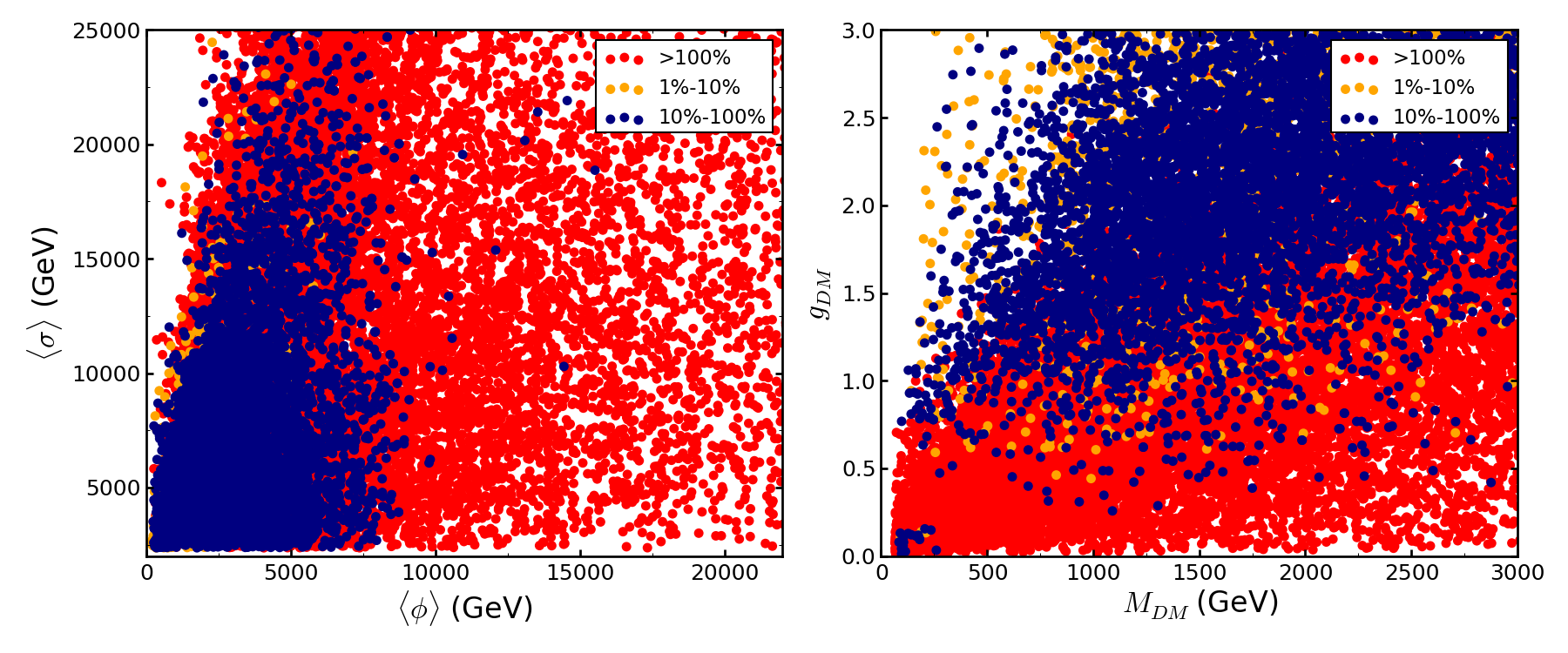}}
\caption{\label{fig:vevsgCW} Left panel: Scatter plot of the vev $\langle 
\phi \rangle$ versus the vev of the scalar singlet $\langle \sigma \rangle$. 
Due to the small mixing angles, 
we can see that the dark matter relic density is almost independent of $\langle \sigma \rangle$. Right panel: Scatter plot of the dark matter mass $M_{Z'}$ versus the gauge coupling $g_{\text{\tiny{DM}}}$. Different colours indicate whether the vector gauge triplet accounts for more or less than 100$\%$, 10$\%$ 
and 1$\%$ of the observed dark matter abundance.
}
\end{figure}

\begin{figure}[tbp]
\centerline\\  \center
\scalebox{0.5}{\includegraphics{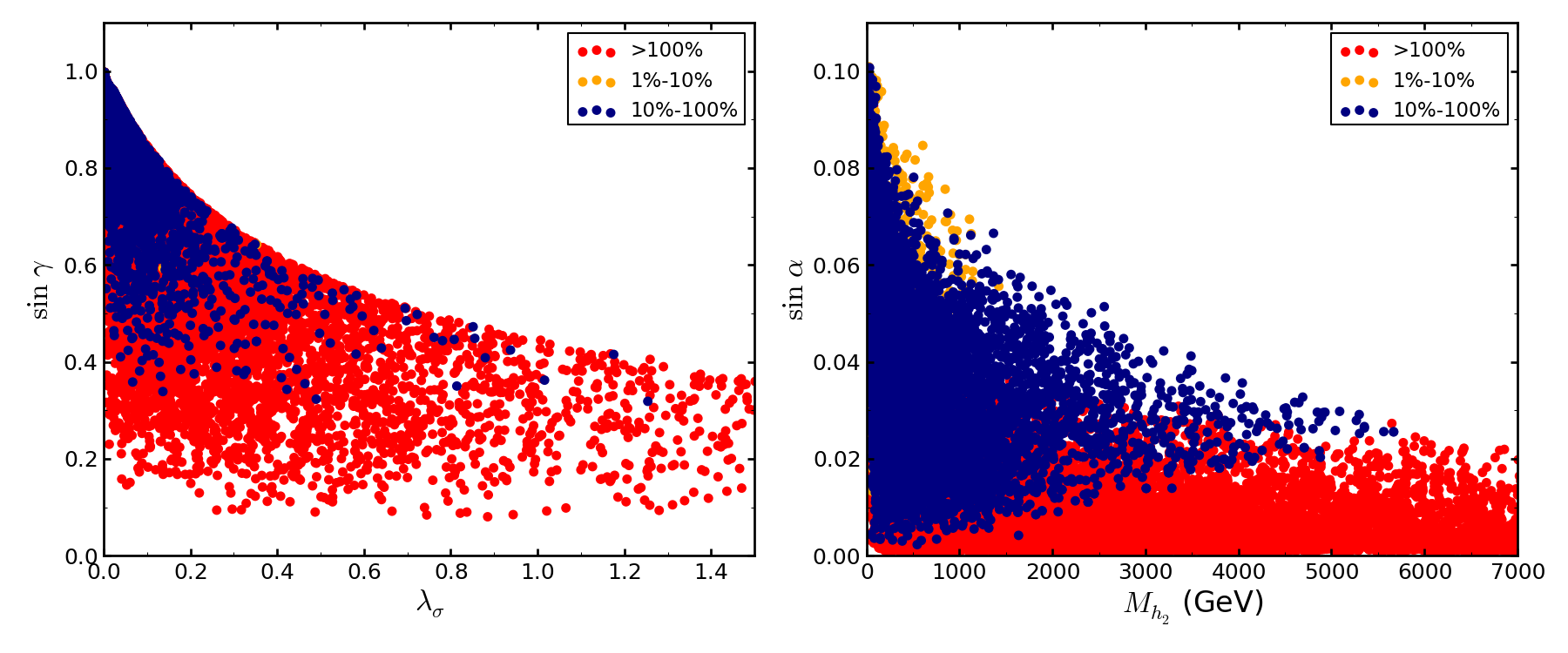}}
\caption{\label{fig:sines} Left panel: Scatter plot of $\sin \gamma$ against the quartic coupling $\lambda_{\sigma}$. Larger values of $\sin \gamma$ are preferred. 
Right panel: Scatter plot of $\sin \alpha$ versus the scalar mass $M_{h_2}$.
Due to $\langle \sigma \rangle \gg \langle h \rangle$ we get small values for the mixing angle $\alpha$. 
Different colours indicate whether the vector gauge triplet accounts for more or less than 100$\%$, 10$\%$ 
and 1$\%$ of the observed dark matter abundance.
}
\end{figure}
After radiative symmetry breaking breaking of SU(2$)_{\text{DM}}$ by $\Phi$, which is in the fundamental representation of the group, there is a remnant SO(3) symmetry that ensures the three gauge bosons $Z'^{\hspace{0.25mm}a}$ acquire the same mass
 $M_{Z'} = \frac{1}{2}\,g_{\text{\tiny{DM}}} \langle \phi \rangle$, and are stable.
In contrast to models where the DM is odd under a $\mathbb{Z}_2$ discrete symmetry, in the present scenario we can have 
dark matter semi-annihilation processes where a DM particle is also present in the final state. The DM annihilation diagrams are shown in 
Figs.~\ref{fig:AnnSM} and \ref{fig:AnnScalars}, while the semi-annihilation ones are shown in Fig.~\ref{fig:SemiDiagrams}.

Also, due to the radiative generation of $\langle \phi \rangle$ in most region of  parameter space the scalar mass will be smaller than the gauge boson mass, $M_{h_2}\!<\!M_{Z'}$. This means that semi-annihilation processes $Z'^{\hspace{0.25mm}a} Z'^{\hspace{0.30mm}b} \rightarrow Z'^{\hspace{0.25mm}c}\hspace{0.5mm}h_i$ will be dominant over annihilation ones in most of the parameter space. To leading order the non-relativistic cross-section from the semi-annihilation diagrams is given by ({\it cf.} \cite{Khoze:2014xha}),

%
\begin{equation}
\label{eq:xsecijk}
\langle\sigma_{abc} v\rangle=\frac{3 g_{\text{\tiny{DM}}}^4}{128 \pi}\frac{(O_{2i})^2}{ 
M_{Z'}^2}\left(1-\frac{M_{h_i}^2}{3 M_{Z'}^2}\right)^{-2}\left(1-\frac{10 
M_{h_i}^2}{9 M_{Z'}^2} +\frac{M_{h_i}^4}{9 M_{Z'}^4}\right)^{3/2}\;.
\end{equation}

In order to take into account all annihilation channels into SM particles and properly take into account thresholds and resonances we have implemented the model in \texttt{micrOMEGAs 4.1.5} \cite{Belanger:2001fz}.
We fix the dark matter relic abundance from the latest Planck satellite measurement $\Omega h^2 = 0.1197 \pm 0.0022$ 
\cite{Ade:2015xua}. 
Figure~\ref{fig:Masses} shows the dark matter fraction as a function of $M_{Z'}$ and the scalar mass $M_{h_2}$; 
the isolated strip of points on the left side of the plots corresponds to the resonance $M_{h_2}\!\approx 2 M_{Z'}$.

On the left plot in Fig.~\ref{fig:Masses} there is a large red coloured region on the left side (producing too much dark matter), in this region $M_{h_2}$ has a close value to $M_{Z'}$ (note that this region does not exist in the Coleman-Weinberg limit). This region exists thanks to very large values of $M_{h_3}$ and $\langle \phi \rangle \gg M_{Z'}$. 
In the left panel of Fig.~\ref{fig:vevsgCW} we show the dark matter fraction as a function of both vevs, $\langle \phi \rangle$ and $\langle \sigma \rangle$, from this plot we see there is an upper bound on $\langle \phi \rangle$ in order not to overproduce dark matter, $\langle \phi \rangle<17$ TeV. Later on we shall see that there is a lower bound on $\langle \sigma \rangle$ coming from leptogenesis, $\langle \sigma \rangle>2.5$ TeV, we have already imposed this bound on all the scatter plots we show.

In the right panel of Fig.~\ref{fig:vevsgCW} we show the dark matter fraction as a function of $M_{Z'}$ and the gauge coupling $g_{\text{\tiny{DM}}}$. In this plot it becomes clear that as we increase the gauge coupling, the relic density decreases. 
The left panel of Fig.~\ref{fig:sines}  shows the same analysis for the mixing angle $\sin \gamma$ and the quartic couplng $\lambda_{\sigma}$. Here we can already notice a preference for the region $\sin \gamma\approx\!1$, where $\lambda_{\sigma}$ takes on small values and $\langle \sigma \rangle\gg \langle \phi \rangle$. Due to the lower bound on $\langle \sigma \rangle$ the mixing angle $\alpha$ takes on very small values, this is shown in the right panel of Fig. 6. 
%
\begin{figure}[tbp]
\centerline\\  \center
\scalebox{0.5}{\includegraphics{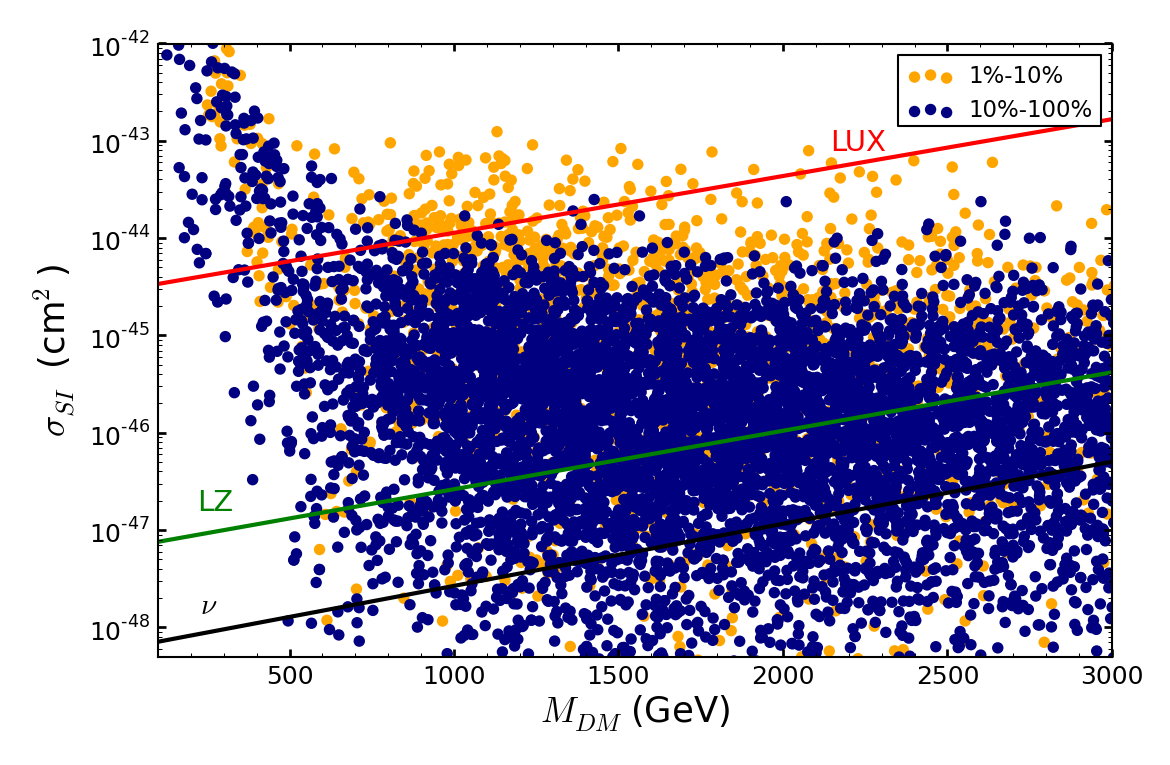}}
\caption{\label{fig:plotDD} Spin-independent DM-nucleon cross section as a 
function of the DM candidate mass $M_{Z'}$. We show current experimental limits 
from LUX \cite{Akerib:2013tjd} (red line), future limits from LZ \cite{Malling:2011va} (green 
line) 
and the neutrino coherent scattering limit \cite{Billard:2013qya} (black 
line).}
\end{figure}

The spin-independent cross section between $Z'^{\hspace{0.25mm}a}$ and a nucleon is given by,
\begin{equation} \label{eq:DDcrossSI}
\sigma_{\text{\tiny SI}} = 
\frac{f_N^2\hspace{0.5mm}m_N^4\hspace{0.5mm}M_{Z'}^2}{\pi\hspace{0.5mm}\langle h 
\rangle^2 \hspace{0.5mm} \langle \phi \rangle^2} \left( \sum_{i=1}^3 
\frac{O_{2i}O_{1i}}{M_{h_i}^2} \right)^2\,,
\end{equation}
%
where $m_N$ is the nucleon mass,  $f_N=0.303$ \cite{DiChiara:2015bua} is the nucleon form-factor, and
$O_{ij}$ are the elements of the rotation matrix Eq.~\eqref{eq:rotationmatrix} that relates the 
scalar mass eigenstates states to the ones in the Lagrangian. 
This orthogonal matrix $O$ is the one that diagonalises the mass matrix. 
Due to the form of this matrix, the direct 
detection diagrams have a destructive interference when the scalar state with a large $\phi$ component has a mass very close to $M_{h_1}$, 
this has been previously noted in \cite{Baek:2012se, Hambye:2013sna}; while the scalar state with a large $\sigma$ component has no 
direct couplings either to dark matter or to Standard Model particles and hence gives 
only a small contribution to $\sigma_{\text{\tiny SI}}$.
%
\begin{figure}[tbp]
\centerline\\  \center
\scalebox{0.8}{\includegraphics{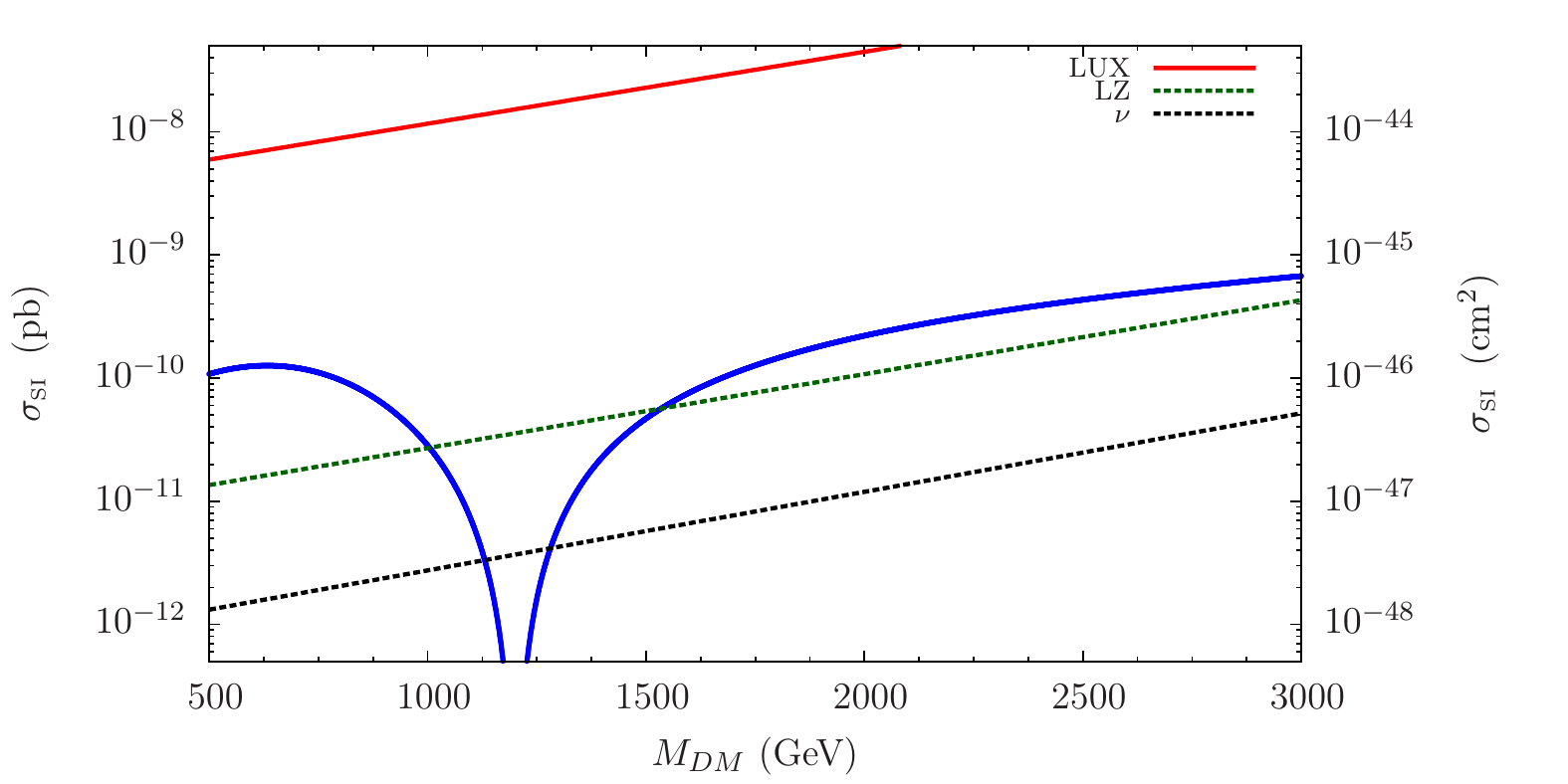}}
\caption{\label{fig:plotDDMA} Spin-independent DM-nucleon cross section as a 
function of the vector DM candidate mass $M_{Z'}$, for benchmark point BP 1. We show current experimental limits 
from LUX \cite{Akerib:2013tjd} (red line), future limits from LZ \cite{Malling:2011va} (green 
line) 
and the neutrino coherent scattering limit \cite{Billard:2013qya} (black 
line). To generate this plot we fix all the scalar couplings and vary only $g_{\text{\tiny{DM}}}$, which means that $M_{Z'}$ and $M_{h_2}$ are also varied
while all other parameters remain fixed. }
\end{figure}
%
Figure~\ref{fig:plotDD} shows that except for resonances, the region with $M_{Z'}\!<\!250$ GeV has been already excluded 
by the existing experiments, while a large region of parameter space will be tested by future underground experiments such as LZ \cite{Malling:2011va} and XENON1T \cite{Aprile:2015uzo}. In Fig.~\ref{fig:plotDDMA} we show the direct-detection cross section as a function of the dark matter mass for benchmark point BP 1, we fix all the scalar couplings and vary only $g_{\text{\tiny{DM}}}$, the dip corresponds to $M_{h_2}\!\approx\!M_{h_1}$.

\medskip

\section{Leptogenesis via oscillations of right-handed neutrinos}
\label{sec:Leptogenesis}

\medskip
Leptogenesis is an attractive and minimal mechanism to solve the baryon 
asymmetry of the universe (BAU). This means being able to produce the observed value of
\begin{equation} \label{eq:DDcrossB}
\frac{n_{b_{\text{obs}}}}{s} = (8.75 \pm 0.23) \times 10^{-11}.
\end{equation}
In the Akhmedov-Rubakov-Smirnov framework \cite{Akhmedov:1998qx} a lepton flavour asymmetry 
is produced during oscillations of the right-handed Majorana neutrinos $N_i$ with 
masses around the electroweak scale or below, which makes this approach 
compatible with classical scale invariance.\footnote{In the sense that no additional very large scales are required 
to be introduced in the model to make this type of leptogenesis work.}
From Big Bang nucleosynthesis we obtain the lower bound $M_N\!>\!200$ MeV, in order not to spoil primordial nucleosynthesis.
For our calculations we make use of the the Casas-Ibarra parametrisation \cite{Casas:2001sr} for the matrix $Y^D$,
\begin{equation}
  \label{eq:CasasIbarra} 
Y^{D\, \dagger} \,=\, U_{\nu} \cdot \sqrt{m_{\nu}} \cdot {\cal R} \cdot \sqrt{M_N} \times \frac{\sqrt{2}}{\langle h \rangle}\,,
\end{equation}
where $m_\nu$ and $M_N$ are diagonal mass matrices of active and Majorana neutrinos respectively.
The active-neutrino-mixing matrix $U_{\nu}$ is the PMNS matrix which contains six real parameters, including three measured mixing angles 
and three CP-phases. The matrix ${\cal R}$ is parametrised by three complex angles $\omega_{ij}$.  
Using this framework with three right-handed neutrinos one can generate the correct baryon asymmetry without requiring tuning the $N_i$ mass splittings, but rather enhancing the entries in the Dirac Yukawa matrix through the imaginary parts of the complex angles $\omega_{ij}$ \cite{Shuve:2014zua}.

Due to the non-trivial topological structure of the vacuum in SU(2$)_L$ there exist electroweak sphaleron processes which violate $B+L$ quantum number,
and these will transfer the lepton flavour asymmetry $n_{Le}$ into a baryon asymmetry $n_{b}$, with the conversion factor given by,
\begin{equation} \label{eq:conversion}
\frac{n_{b}}{s} \,\simeq\, -\frac{3}{14} \times 0.35 \times \frac{n_{Le}}{s}.
\end{equation}
%
A critical condition for the mechanism of \cite{Akhmedov:1998qx} to work, 
is that two of three neutrino flavours, $N_2$ and $N_3$, should come into thermal equilibrium 
with their Standard Model counterparts before the universe cools down to $T_{\text{\tiny EW}}$ (when electroweak sphaleron processes freeze out), while the remaining flavour does not. In other words, the present mechanism consists of different time scales $T_{\rm osc}\gg T_{eq_3} \sim T_{eq_2} > T_{\text{\tiny EW}} > T_{eq_1}$, where $T_{eq_i}$ represents the temperature at which $N_i$ equilibrates with the thermal plasma and $T_{\rm osc}$ is the temperature at which the oscillations start to occur.
In terms of the decay rates for the three sterile neutrino flavours this implies,
\begin{equation}
 \Gamma_2(T_{\text{\tiny EW}}) \,>\, H(T_{\text{\tiny EW}}) \ , \quad 
  \Gamma_3(T_{\text{\tiny EW}}) \,>\, H(T_{\text{\tiny EW}}) \ , \quad \Gamma_1(T_{\text{\tiny EW}}) \,<\, H(T_{\text{\tiny EW}}),
\end{equation}  
where $H$ is the Hubble constant, 
\begin{equation}
 H(T) \,=\, \frac{T^2}{M_{\rm P}^*} \, , \qquad 
 M_{\rm P}^*\,\equiv \, \frac{M_{\rm P}}{\sqrt{g_*}\sqrt{4\pi^3/45}} \,\simeq\, 10^{18}\, {\rm GeV}
  \end{equation} 
 and $M_{\rm P}^*$ is the reduced Planck mass.
Therefore, we require,
\begin{equation}
  \label{eq:washout}
\Gamma_1(T_{\text{\tiny EW}})\,=\, \frac{1}{2}\sum_i
Y^{\rm D \,\dagger}_{ei}  Y^{\rm D }_{ie} \,\gamma_{av} \, T_{\text{\tiny EW}}\, <\, H(T_{\text{\tiny EW}}) \,.
 \end{equation}
Here the dimensionless quantities $\gamma_{av} \approx 3 \times 10^{-3}$
are derived from the decay rates of the right-handed neutrino $N_e$ of the `electron flavour' 
 tabulated in Ref.~\cite{Besak:2012qm}. These right-handed neutrino decay (or equivalently production) rates were computed  
 in~\cite{Besak:2012qm} using 
 $1 \leftrightarrow 2$ and $2 \leftrightarrow 2$ processes\footnote{These processes are shown in Figs. 1 and 2 in  Ref.~\cite{Besak:2012qm} and contain a single external $N$ leg -- as relevant for the $N$-production or decay processes of interest.}
 involving the neutrino vertices
 $Y_{ai}^{D\dagger} \hspace{0.5mm} \overline{l_{La}} (\varepsilon H)^{\dagger} N_i$
  and 
  $ Y_{ia}^D \hspace{0.5mm} \overline{N_i} (\varepsilon H) l_{La}$
  with the Dirac Yukawas.

One can also ask if the new interactions present in our model, those involving the Majorana Yukawas, $\frac{1}{2} \,Y_{ij}^M \sigma \overline{N_i}^c N_j$
and $\frac{1}{2} \,Y_{ij}^{M\dagger} \sigma \overline{N_i} N_j^c$,
could affect the dynamics. These interactions always contain a pair of right-handed neutrinos and do not change the right-handed neutrino number (the singlet $\sigma$ carries the $N$-number $-2$ but above the electroweak phase transition temperature, the vev of $\sigma$ vanishes). Hence these processes could contribute to the $N$ production or decay into the Standard Model particles only in combination with other interactions.
As the Majorana Yukawa couplings are small $Y^M\!\approx\! 10^{-5}$ on the part of the parameter space relevant for us (see Table \ref{tab:2}) and the cross-section being proportional to $(Y^M)^2$ means that these interactions will give subleading effects to all the processes considered in \cite{Besak:2012qm}. Therefore, we can follow \cite{Drewes:2012ma} and make the assumption that the number density of sterile neutrinos is very small compared to their equilibrium density at high temperatures, $T_{\rm osc}\approx 10^6$ GeV, around which the main contributions to the lepton-flavour asymmetry are generated. 

\begin{figure}[tbp]
\centerline\\  \center
\scalebox{0.45}{\includegraphics{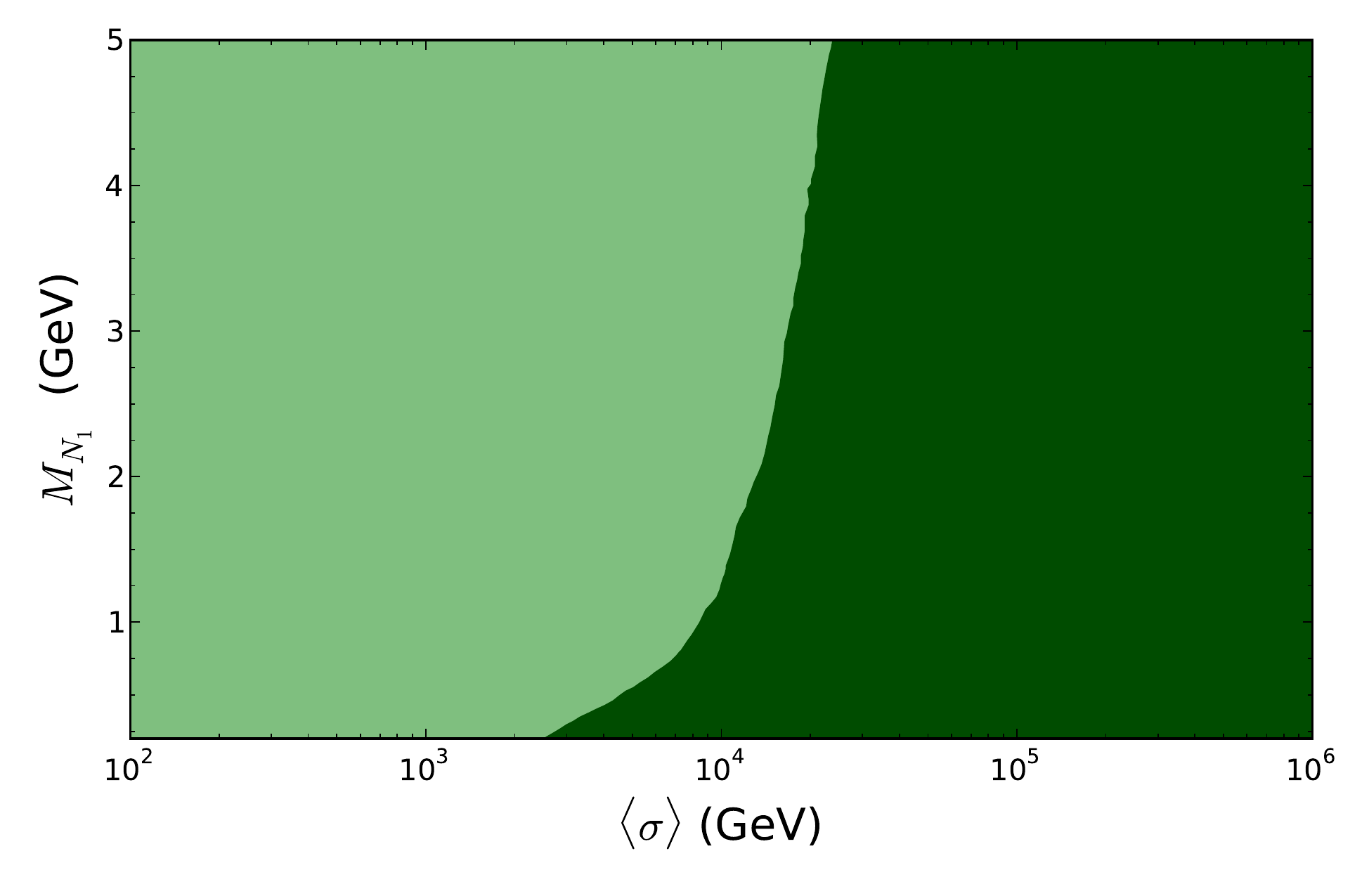}}
\caption{\label{fig:massLimit} The region in dark green can explain the baryon asymmetry through leptogenesis; we have fixed the 
mass splittings to be $\Delta M_{N_i}\!\geq\!M_{N_1}/10$. This 
plot shows that there is a lower bound $\langle \sigma \rangle > 2.5$ TeV in order to
produce the correct amount of baryon asymmetry. The region in light green cannot produce enough baryon symmetry and/or does not satisfy the wash-out criterion Eq.~\eqref{eq:washout}.}
\end{figure}

It was already shown in~\cite{Khoze:2013oga} that flavoured leptogenesis 
can work in a classically scale invariant framework. In their set-up three right-handed neutrinos 
are coupled to a scalar field that acquires a vev, as in the present model. The main difference being that in the present scenario we have not gauged the $B\!-\!L$ quantum number. We quote the final result for the lepton flavour asymmetry 
(of $a$th flavour) obtained in ~\cite{Khoze:2013oga} from extending the results of Ref.~\cite{Drewes:2012ma} 
to the classically scale-invariant case,
\begin{equation}
  \label{eq:deltaLOUR}
\frac{n_{La}}{s} \,=\, - \gamma_{av} ^2 \times 7.3 \times 10^{-4} \sum_{c}\sum_{i\neq j} i \, 
 (Y^{\rm D \,\dagger}_{ai} Y^{\rm D}_{ic} Y^{\rm D \,\dagger}_{cj} Y^{\rm D }_{ja} -
 Y^{\rm D \,t}_{ai} Y^{\rm D\, *}_{ic} Y^{\rm D \,t}_{cj} Y^{\rm D \, *}_{ja})\, \times {\cal I}_{ij}\, ,
 \end{equation}
where the quantity ${\cal I}_{ij}$ is given by,
\begin{equation}
  \label{eq:Iij} 
 {\cal I}_{ij}\,=\, 
  \frac{16}{ \sum_k (Y^{\rm M \,\dagger}_{ik} Y^{\rm M}_{ki} -Y^{\rm M \,\dagger}_{jk} Y^{\rm M}_{kj})}\,
  \frac{M_{\rm P}}{\langle \sigma \rangle}\,
  \left( 1 -  \frac{\langle \sigma \rangle}{T_{\rm osc}} +\frac{1}{4} {\rm tan}^{-1}\left(\frac{4 \hspace{0.5mm} \langle \sigma \rangle}{T_{\text{\tiny EW}}}\right) -
 \frac{1}{4} {\rm tan}^{-1}\left(4\right) 
  \right)\,,
 \end{equation}
for $\langle \sigma \rangle\, <\, T_{\rm osc}$. For the case $\langle \sigma \rangle\, \geq \, T_{\rm osc}$ and further details on the derivation of Eq.~\eqref{eq:deltaLOUR} we refer the reader to Ref.~\cite{Khoze:2013oga}.
It follows from \eqref{eq:Iij} that the amount of the
lepton flavour asymmetry  is proportional to $\langle \sigma \rangle\, M_P / \Delta M_{N_i}^2$.
Hence if we want to avoid any excessive fine-tuning of the mass splittings between different flavours of Majorana neutrinos,
the relatively large values of $\langle \sigma \rangle \gtrsim 10^4$ GeV are preferred. 
From Fig.~\ref{fig:massLimit} we can see that there is a lower 
bound on $\langle \sigma \rangle$ if we impose some restriction on the mass splittings of the
right-handed neutrinos. In view that we would like to stay far away from the fine-tuning region,
we impose $\Delta M_{N_i}\!\geq\!M_{N_1}/10$ which gives the limit 
$\langle \sigma \rangle > 2.5 $ TeV in order for leptogenesis to explain the baryon asymmetry.
Imposing this condition removes the points with very small mixing angle $\gamma$, as can be seen in the left panel of Fig.~\ref{fig:sines}. 

As we can see from Fig.~\ref{fig:massLimit} there is also an upper bound on $M_{N_i}$ for each value of $\langle \sigma \rangle$, this bound is mainly due to the wash-out criterion Eq.~\eqref{eq:washout} not being satisfied any more. This upper bound becomes weaker once we reach $\langle \sigma \rangle \geq 10^4$ GeV. 
This sits well with our approach based on the common dynamical origin of all vevs:
 once an explanation for dark matter is included, 
$\langle \sigma \rangle$ cannot be too large compared to $\langle \phi\rangle$.

The procedure to obtain the plot in Fig.~\ref{fig:massLimit} is as follows. We fix the complex phases $\omega_{12}$ and $\omega_{13}$ to the benchmark values given in \cite{Drewes:2012ma} ($\omega_{12}=1+2.6i$ and $\omega_{13}=0.9+2.7i$), and for each point we scan over $\omega_{23}$, if we find at least one point that works well then we label it as a good point (dark green) otherwise it is a bad point (light green). In further scans we have found that varying $\omega_{12}$ and $\omega_{13}$ has a negligible impact on the final results.

The generated total lepton asymmetry is proportional to $\langle\sigma\rangle$, ({\it cf.}~\eqref{eq:deltaLOUR}, \eqref{eq:Iij})
\begin{equation} \label{eq:DDcrossL}
n_L \sim (Y^D)^4\, \frac{ \langle \sigma \rangle\,  M_P}{\Delta M_{N_i}^2} \sim\, \langle\sigma\rangle\,  M_P \, 
\frac{m_{\nu}^2}{v^4}\,,
\end{equation}
where we used the see-saw mechanism for the masses $m_{\nu}$ of visible neutrinos, and  $v$ is the SM Higgs vev.
Hence $n_L$ vanishes as $\langle\sigma\rangle$  approaches zero. This also explains why in Fig.~\ref{fig:massLimit}, there is a stronger dependence on $\langle \sigma \rangle$ than on the masses $M_{N_i}$.

\medskip

We carried out a scan over all free parameters in our model to determine the region of the parameter space
where the leptogenesis mechanism outlined above can generate the observed baryon asymmetry. At the same time
we require that the model provides a viable candidate for cosmological dark matter. 
We would like to mention in passing that all the present results on leptogenesis also hold when a generic scalar generates a mass for the sterile neutrinos (i.e with no reference to classical scale invariance).

The results of the scan and the connection between 
the leptogenesis and dark matter scales are reviewed in the following Section. Furthermore, in Tables \ref{tab:1}  and \ref{tab:2} 
we present four benchmark points to illustrate the viable model parameters.
In the remainder of this Section we would like to comment on the choice of parameters for the leptogenesis part of the story.

We first note that our leptogenesis realisation does not require any sizeable fine-tuning of the mass splittings $\Delta M_{N_i}$. 
 For example our first
 benchmark point BP 1 has ({\it cf.} Table~\ref{tab:2}),
 \begin{equation}
 \label{eq:MNBP1}
 M_N = ( 0.225\,,\, 0.25\,,\, 0.275) \, {\rm GeV}.
 \end{equation}
 At the same time, the masses of active neutrinos are set to agree with the observed mass splittings; for BP 1 we have,
  \begin{equation}
 \label{eq:MnuBP1}
 m_{\nu} = ( 0\,,\, 8.7\,,\, 49.0) \, {\rm meV}.
 \end{equation}
 The lepton asymmetry \eqref{eq:deltaLOUR} also depends on the matrix of Dirac Yukawa couplings $Y^D$.
 We compute  $Y^D$ in the
 Casas-Ibarra parametrisation  Eq.~\eqref{eq:CasasIbarra} using~\eqref{eq:MNBP1} and \eqref{eq:MnuBP1}
 along with the PMNS matrix and the ${\cal R}$ matrix.
 We have carried out  a general scan on the complex angles $\omega_{ij}$ of the ${\cal R}$ matrix 
 and found that having non-vanishing 
 $\text{Im}[\omega_{ij}]$ is important in order to obtain the required amount of lepton asymmetry.\footnote{Note that positive values of $\text{Im}[\omega_{ij}]$ enhance the elements of the Dirac Yukawa matrix $Y^D$.}
At the same time this does not lead to any excessive fine-tuning.  We have checked this for the 
numerical values of ${\cal R}$ matrix elements 
in our scan. For example, for
BP 1 we have (using the $\omega_{ij}$ values in Table  \ref{tab:2}),
\begin{equation}\label{eq:Rmatrix}  
{\cal R} =  \begin{pmatrix}
 -36.52-33.80 i \hspace{4mm}& 34.11\, -36.97 i \hspace{4mm}& 5.854\, +4.604 i \\
 84.43\, +100.0 i \hspace{4mm}& -101.0+85.98 i \hspace{4mm}& -16.63-14.20 i \\
 -105.4+91.81 i \hspace{4mm}& -93.42-106.4 i \hspace{4mm}& 14.94\, -17.61 i \\
 \end{pmatrix},
\end{equation}
and the resulting  matrix of Dirac Yukawa couplings, 
\begin{equation}\label{eq:YDmatrix}  
Y^D =  \begin{pmatrix}
 17.87\, -2.12i \hspace{4mm}  & -73.37-125.6 i \hspace{4mm}   & -210.9-127.3 i \\
 -2.168-19.11 i \hspace{4mm}  & -134.4+77.79 i \hspace{4mm}   & -136.9+224.6 i \\
 -3.395-0.2434 i \hspace{4mm} & 9.677\, +24.56 i \hspace{4mm} & 34.69\, +28.93 i \\
 \end{pmatrix}\times 10^{-8}.
\end{equation}
These matrices do not exhibit a high degree of tuning, and we have checked that this is also the case for generic points of our scan.

\medskip

\section{Connection among the scales}
\label{sec:connection}

After having performed a scan over all free parameters in our model, we find that:
\newline
(1) $\quad \langle \phi \rangle < 17$ TeV in order for dark matter not to overclose the universe, and 
\newline
(2) $\quad \langle \sigma \rangle > 2.5$ TeV in order in order for leptogenesis to explain the baryon asymmetry. 

\medskip

\noindent From the left plot of Fig.~\ref{fig:sines} we can see that the interesting region in parameter space has large values of $\sin \gamma$, and with this in mind we can separate the interesting regime into two regions:

\begin{enumerate}
 \item  $\langle \sigma \rangle \approx \langle \phi \rangle \sim$ TeV\\
In this region\footnote{Recall that 
$\tan^2 \gamma= \langle\sigma\rangle^2/\langle \phi\rangle^2$.}
 we have $\sin \gamma \approx\cos \gamma$ $(\gamma\approx\pi/4)$ so there is a strong mixing between the scalar states 
$\phi$ and $\sigma$, and due to the Gildener-Weinberg conditions $\lambda_{\phi}\approx\lambda_{\sigma}$. To avoid overproducing DM, both $\langle \sigma \rangle$ and $\langle \phi \rangle$ have to be less than 10 TeV. Due to the not so large values of $\langle \sigma \rangle$, a large part of this region requires some amount of fine-tuning of the right-handed neutrino mass splittings in order for leptogenesis to work. The use of the Gildener-Weinberg mechanism is crucial in this region.
\item   $\langle \sigma \rangle \gg \langle \phi \rangle \sim$ TeV\\
In this region we have $\sin \gamma \approx 1$, so it can be seen as the Coleman-Weinberg limit of the more general Gildener-Weinberg mechanism. The scalar $\sigma$ overlaps maximally with $h_2$ and can be thought of as the Coleman-Weinberg scalar. In this region the radiative symmetry breaking is induced by $\lambda_{\sigma}\ll1$ and we get $M_{h_2}\!\ll\!M_{h_3}$. This region also corresponds to the majority of good (blue) points in Figs.~\ref{fig:Masses}-\ref{fig:sines}.
Most points have $M_{\text{DM}}>M_{h_2}$. This is the region of most interest since the large values of $\langle \sigma \rangle$ require almost no fine-tuning in $\Delta M_{N_i}$ in order for leptogenesis to work. 
\end{enumerate}
\begin{table}[t]
\centering
\begin{tabular}{|l| c| c| c| c| }
\hline
& BP 1  & BP 2 & BP 3 & BP 4 \\ \hline
$\Omega h^2$                   & $0.122$                      & $0.12$                          & $0.12$                   & $0.118$                 \\
$\sigma_{\text{\tiny{SI}}}$ (cm$^2$)  & $1.90 \times 10^{-46}$     & $3.32 \times 10^{-46}$      & $1.06 \times 10^{-46}$  & $3.11 \times 10^{-47}$   \\
$\langle h \rangle$ (GeV)             & $246$                      & $246$                       & $246$                   & $246$ \\
$\langle \phi \rangle$ (GeV)  & 2260       & 1260    &  1020  &  4590    \\
$\langle \sigma \rangle$ (GeV)  & 3080    & 5930       & 2830 & 11790   \\
\hline
$\lambda_{h\phi}$   &        0.035      &     0.406      & -0.335    & 0.017\\
$\lambda_{\phi \sigma}$  &   0.164       &      0.122        & 0.40     & 0.141\\
$\lambda_{h\sigma}$    &     0.0185       &     0.018       & -0.045    & 0.003\\
$\lambda_h$   &              0.131       &       0.159     & 0.147     & 0.130 \\
$\lambda_{\sigma}$&          0.044       &      0.003    & 0.027    & 0.011\\
$\lambda_{\phi}$&            0.152    &       1.352      & 1.527     & 0.464\\
\hline
$g_{\text{\tiny{DM}}}$   &0.61          &  1.39       &0.96  &2.41 \\
$M_{h_1}$ (GeV)   & 125     & 125          & 125       & 125  \\
$M_{h_2}$ (GeV)   & 81.6     & 94.1       & 137.3      & 839.1\\
$M_{h_3}$ (GeV)   & 1544   & 2124      & 1900          & 4745  \\
$M_{Z'}$ (GeV)    &690      & 880      & 490         & 5527  \\ 
\hline
$\sin \alpha$& 0.06     &  0.04    & 0.08    & 0.02 \\
$\sin \beta$&  0.01     &  0.03     &  -0.025   &  0.001\\
$\sin \gamma$&  0.80     &  0.98       & 0.94    & 0.93 \\
$\mu_{\text{\tiny{GW}}}$ (GeV) &   829   &  1149   & 1110 & 4550  \\
\hline
\end{tabular}
\caption{
  \label{tab:1}
Four benchmark points for the model presented in this work. All four points give the correct dark matter abundance within $2\sigma$.
}
\end{table}

\noindent In Table \ref{tab:1} we give a set of benchmark points that satisfy all experimental 
constraints and give the correct dark matter abundance within $2\sigma$. The benchmark points BP1, BP2 and BP3 are within 
reach of future direct detection dark matter experiments.
For these same points we provide in Table \ref{tab:2} numerical values that generate the correct amount of baryon asymmetry via leptogenesis.
We work with the current experimental central values for the neutrino sector taken from \cite{Gonzalez-Garcia:2014bfa}, we assume normal ordering for the active neutrino masses. The values for $\langle Y^D\rangle$ are computed as the average of $\sqrt{2M_Nm_{\nu}}/\langle h \rangle$. This estimate corresponds to the naive see-saw relation and it is smaller than the actual entries in the matrix $Y^D$ due to the enhancement by the imaginary parts of $\omega_{ij}$ in the ${\cal R}$ matrix. Nevertheless, for our benchmark points these enhancement factors are always less than $1.5\times 10^2$.

\begin{table}[t!]
\centering
\begin{tabular}{|l| c| c| c| c|}
\hline
& BP 1  & BP 2 & BP 3 & BP 4 \\ \hline
$\left<\sigma\right>\hspace{2mm}$  (GeV)   & 3080 &     5930        &  2830 & 11790 \\
$M_{N_1}$ (GeV)               & $0.225$ &    $0.30$   & $0.20$   & $0.9$\\
$M_{N_2}$ (GeV)               & $0.25$ &     $0.33$   & $0.22$  & $1.0$ \\
$M_{N_3}$ (GeV)               & $0.275$ &    $0.36$   & $0.24$   & $1.1$ \\
\hline
$m_1$ (meV) & $0.0$ & $0.0$    & $0.0$    & $0.0$ \\
$m_2$ (meV) & $8.7$ & $8.7$    & $8.7$    & $8.7$\\
$m_3$ (meV) & $49.0$ & $49.0$  & $49.0$ & $49.0$\\
$\sin \theta_{12}$     & 0.55 & 0.55 & 0.55 & 0.55 \\
$\sin \theta_{23}$     & 0.67 & 0.67 & 0.67 & 0.67\\
$\sin \theta_{13}$     & 0.15 & 0.15 & 0.15 & 0.15\\
$\delta$              & $-\pi/4$   & $-0.6$    &$ -\pi/4$     &$ \pi$ \\
$\alpha_1$            & 0          & 0.3           & 0            & $-\pi$ \\ 
$\alpha_2$            &$-\pi/2$    & $-1.1$    &$ -\pi/2$     &$ \pi$ \\
$\omega_{12}$    &    $1.5+2.6i$         &    $1.5+2.6i$      & $1.0+2.6i$      & $1.5+2.6i$ \\
$\omega_{13}$    &     $0.9+2.7i$      &      $0.9+2.7i$      & $0.9+2.7i$       & $0.9+2.7i$\\
$\omega_{23}$    &     $0.03-1.8i$       &     $-0.30-1.4i$      & $0.05-1.85i$  & $-1.4i$\\
\hline
$n_{Le}/(s\times 2.5 \times 10^{-10})$     &  -4.71     &    -5.75     & -5.36   & -6.43\\
$n_{L\mu}/(s\times 2.5 \times 10^{-10})$   &  -1.66    &    -44.18      & 19.03   & -75.82 \\
$n_{L\tau}/(s\times 2.5 \times 10^{-10})$  &  6.37     &     49.93      & -13.67  & 82.25 \\
$\Gamma_e/H(T_{\text{\tiny EW}})$     &    0.90    &    0.82        & 0.91     & 0.98\\
$\Gamma_\mu/H(T_{\text{\tiny EW}})$   &   58.43      &    42.29          & 56.61   & 315.5 \\
$\Gamma_\tau/H(T_{\text{\tiny EW}})$  &    167.63   &    99.03         & 163.07   & 115.56\\
$T_{\text{\scriptsize osc}}$ (GeV)           &   $4.43 \times 10^6$     & $1.90\times 10^6$   & $3.71\times 10^6$ & $4.84\times 10^6$\\
\hline
$Y^{\rm M}_1$              & $\hspace{1mm}7.3\times 10^{-5}\hspace{1mm}$     &$\hspace{1mm}  5.1\times 10^{-5}\hspace{1mm}$                                                                 &         $\hspace{1mm}7.1\times 10^{-5}\hspace{1mm}$  & $\hspace{1mm}7.6\times 10^{-5}\hspace{1mm}$ \\
$Y^{\rm M}_2$              &$ 8.1\times 10^{-5}$     &$  5.6\times 10^{-5}$    &  $7.8\times 10^{-5}$ &  $8.5\times 10^{-5}$ \\
$Y^{\rm M}_3$              & $8.9\times 10^{-5}$     &$  6.1\times 10^{-5}$    &  $8.5\times 10^{-5}$ &  $9.4\times 10^{-5}$ \\
\hline
$\left<Y^{\rm D}\right>$   &  $1.26\times10^{-8}$    &   $1.45\times10^{-8}$      &  $1.18\times10^{-8}$  &  $2.5\times10^{-8}$ \\
\hline
\end{tabular}
\caption{
  \label{tab:2}
Parameters for leptogenesis, same benchmark points as in Table \ref{tab:1}.
}
\end{table}

\begin{figure}[tbp]
\centerline\\  \center
\scalebox{0.45}{\includegraphics{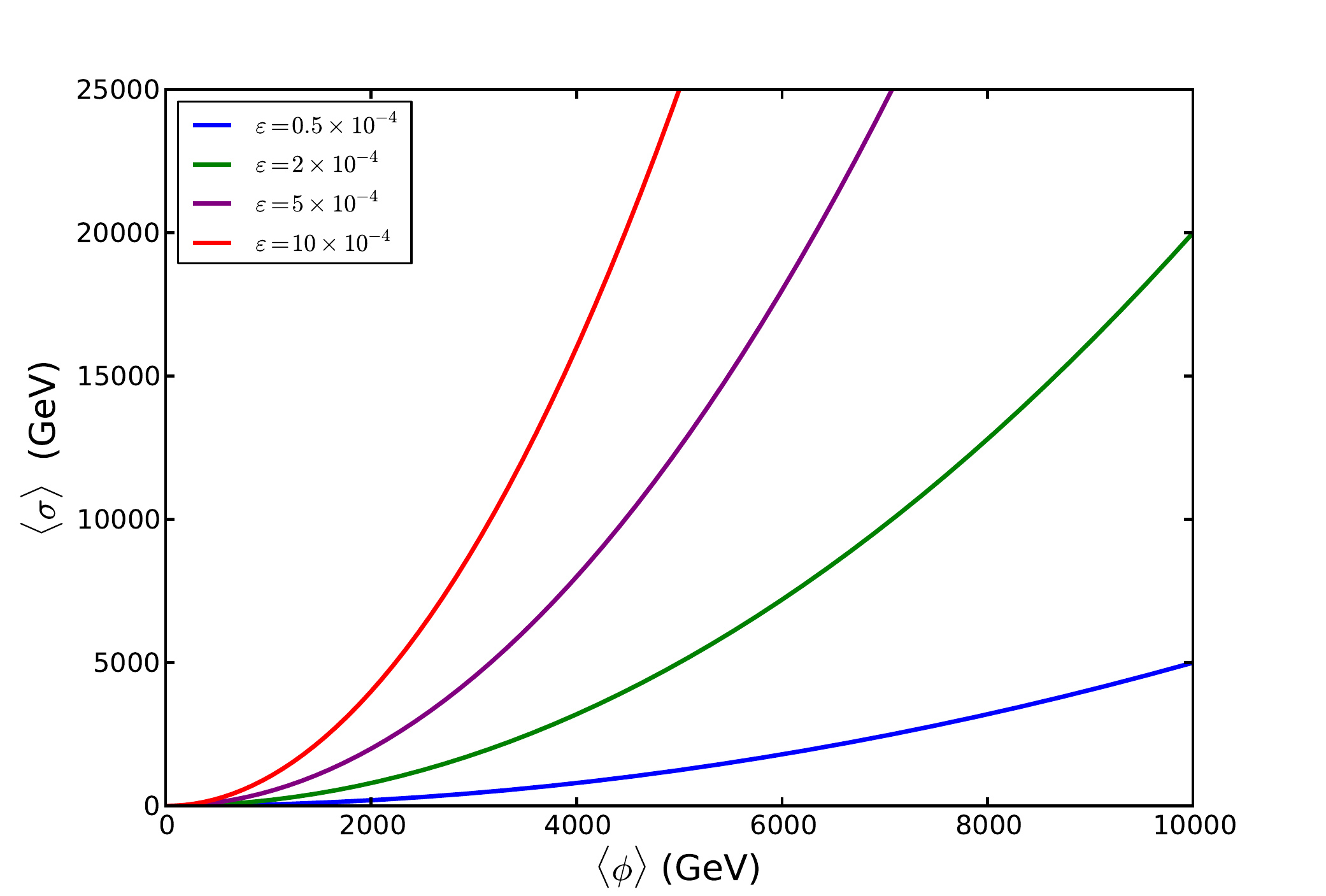}}
\caption{\label{fig:ConnectionVevs} Relation among the two vacuum expectation values, $\langle \phi \rangle$ and $\langle \sigma \rangle$, that yields the observed value of $\Omega_{\text{\tiny{DM}}}h^2 / \Omega_b h^2  = 5.$ Different colours correspond to different values of the parameter $\varepsilon$ defined in Eq.~\eqref{eq:parameterepsilon}.}
\end{figure}

Finding a connection between the scale $\langle \phi \rangle$, responsible for dark matter, and the scale $\langle \sigma \rangle$, responsible for leptogenesis, would be of high interest. From Eq.~\eqref{eq:deltaLOUR} and applying the conversion factor \eqref{eq:conversion}, we can approximate the baryon relic abundance as,
\begin{equation}
\label{eq:Baryon}
\Omega_b h^2 \approx 2.045 \hspace{0.5mm} M_\text{P} \frac{\Delta(Y_D^{4}) \hspace{0.3mm}  \langle \sigma \rangle}{\Delta(M_N^2)}\,.
\end{equation}
%
Regarding the dark matter relic density, in a large portion of our parameter scan semi-annihilations are dominant over annihilations, and hence we can approximate by,
\begin{equation}
\label{eq:DMrelicapprox}
\Omega_{\text{\tiny{DM}}}h^2 \approx \frac{1.07 \times 10^{9} \hspace{0.5mm} x_f}
{\sqrt{g_{\star}} \hspace{0.5mm} M_\text{P} \hspace{1mm} 2\langle\sigma_{abc} v\rangle/3} \hspace{0.5mm} \times \text{GeV}^{-1}\,,
\end{equation}
where $x_f\!=\!M_{Z'}/T_f$, $T_f$ is the freeze-out temperature for dark matter, and $g_{\star}$ is the effective number of relativistic degrees of freedom. A good approximation for the mixing angles is to take $\alpha\!\approx\!\beta\!\approx\!0$ and $\sin\gamma\approx0.9$, substituting these values into Eq.~\eqref{eq:xsecijk} leads to,
\begin{equation}
\label{eq:DMrelic}
\Omega_{\text{\tiny{DM}}}h^2 \approx \frac{7.76 \times 10^{11}}{M_\text{P}} \frac{\langle \phi \rangle ^2}{g_{\text{\tiny{DM}}}^2} \hspace{0.5mm} \times \text{GeV}^{-1}\,.
\end{equation}
%
%
Using Eqs.\eqref{eq:Baryon} and \eqref{eq:DMrelic} we can find the ratio
\begin{equation}
\label{eq:Ratio}
\frac{  \Omega_{\text{\tiny{DM}}}h^2}  {\Omega_b h^2 } \approx \frac{3.79\times 10^{11} \hspace{0.3mm} \Delta(M_N^2) }{M_{\text{P}}^2
\hspace{0.5mm}   g_{\text{\tiny{DM}}}^2 \hspace{0.3mm}  \Delta(Y_D^{4})} \hspace{0.3mm}  \frac{\langle\phi\rangle^2}{\langle\sigma\rangle}
\times \text{GeV}^{-1} = \, 5 \,,
\end{equation}
%
where the last equality comes from the observed relic densities \cite{Ade:2015xua}. After imposing this relation we find a connection
among the scales in the model,
\begin{equation}
\label{eq:Scalesconnection}
\langle \sigma \rangle \,\approx\, \varepsilon \hspace{0.5mm}  
\langle \phi \rangle ^2  \hspace{0.5mm} \times \text{GeV}^{-1}\,, 
\end{equation}
%
where the parameter $\varepsilon$ is defined as,
\begin{equation}
\label{eq:parameterepsilon}
\varepsilon \,= \, \frac{7.59 \times 10^{10} \hspace{0.5mm} \Delta(M_N^2)   }{ M_{\text{P}}^2 \hspace{0.5mm} g_{\text{\tiny{DM}}}^2 \hspace{0.3mm} \Delta(Y_D^{4}) }\, .
\end{equation}
%
The parameter $M_N$ has a dependence on $\langle \sigma \rangle$, but from a physical perspective it is more relevant to fix the mass splittings
rather than the Majorana Yukawa couplings. The parameter $\varepsilon$ gives the connection between both scales, typical values for this parameter are around $10^{-4}$.
Figure~\ref{fig:ConnectionVevs} illustrates this connection between the scales keeping the parameter $\varepsilon$ fixed to different values.
%
%

\medskip

\section{Conclusions}
\label{sec:conclusions}

\medskip

We have presented a model that can explain dark matter and the baryon asymmetry 
of the universe simultaneously, where all the scales in the theory are 
dynamically generated and have a common origin.

In order to ensure the stability of the dark matter candidate, one usually needs to introduce a discrete symmetry by hand. One of the attractive features of the present model is that it leads to a stable DM candidate without the need of introducing an extra discrete symmetry. We already know that in the Standard Model lepton number and baryon number are accidental symmetries, the latter being responsible for the stability of the proton.
In our framework the hidden vector DM is stable due to the accidental non-Abelian global symmetry SO(3).
This accidental symmetry could be broken by non-renormalizable operators leading to the decay of $Z'^{\hspace{0.25mm}a}$ and producing an intense gamma-ray line that could be detected in future experiments \cite{Arina:2009uq}.

The theory also predicts two extra scalar states that have a Higgs-like behaviour and masses around the electroweak scale. From the relation for $\tan^2 \alpha$, Eq.~\eqref{eq:MixAngles}, the interesting region $\langle \sigma \rangle \gg \langle h \rangle$ already requires a small mixing angle $\alpha$ with the SM Higgs boson, due to the small mixing angles we obtain values of $\cos^2 \alpha \cos^2 \beta > 0.95$, so their detection would only be feasible at future colliders. Nevertheless, the LHC at high luminosity will improve the current constraints on the mixing angles $\alpha$ and $\beta$.

From dark matter considerations the value of $\langle \phi \rangle$ is required to be around the TeV scale and due to the common origin of all the vevs, $\langle \sigma \rangle$ cannot be too large, compared to $\langle \phi \rangle$, which means that sterile neutrinos should have small masses of order $\mathcal{O}(1)$ GeV in order for leptogenesis to work without severe tuning of the mass splittings $\Delta M_{N_i}$.
Under some mild assumptions, we found a connection among the scales $\langle \phi \rangle$  (responsible for dark matter) and $\langle \sigma \rangle$ (responsible for leptogenesis) Eq.~\eqref{eq:Scalesconnection}, in order to match the observed ratio $\Omega_{\text{\tiny{DM}}}h^2 / \Omega_b h^2  = 5$.
Using classical scale invariance as an underlying symmetry, we have constructed a minimal extension of the SM that addresses dark matter, the baryon asymmetry of the universe and the origin of the electroweak scale.

\acknowledgments
ADP would like to thank Brian Shuve, Jessica Turner and Ye-Ling Zhou for helpful discussions on the topic of leptogenesis. This work is supported by STFC through the IPPP grant.
ADP acknowledges financial support from CONACyT. 
Research of VVK is supported in part by a Royal Society Wolfson Research Merit Award. 

\bibliography{SU2singlet}


\end{document}